\documentclass[12pt]{article}
\usepackage[T2A]{fontenc}
\usepackage[utf8]{inputenc}
\usepackage{amsthm,amsfonts,amssymb,amscd,cite}
\usepackage[english]{babel}
\usepackage[tbtags]{amsmath}
\usepackage[pdftex]{graphicx}
\usepackage{graphicx}

\let\cal\mathcal

\DeclareSymbolFont{bletters}{OML}{cmm}{bx}{it}
\DeclareMathSymbol{\blad}{\mathord}{bletters}{'025}
\DeclareMathSymbol{\bla}{\mathord}{bletters}{'013}
\DeclareMathSymbol{\bmu}{\mathord}{bletters}{'026}
\DeclareMathSymbol{\bnu}{\mathord}{bletters}{'027}
\DeclareMathSymbol{\bth}{\mathord}{bletters}{'022}
\DeclareMathSymbol{\bfI}{\mathord}{bletters}{"49}
\DeclareMathSymbol{\bdl}{\mathord}{bletters}{"0E}
\DeclareMathSymbol{\bDl}{\mathord}{bletters}{"001}
\def \bpi{\boldsymbol\pi}
\def \bphi{\boldsymbol\phi}

\begin{document}
\title{
$$
{}
$$
{\bf Spin correlation functions, Ramus-like identities, and enumeration of constrained lattice walks and plane partitions}}
\author{
{\bf\Large C.~Malyshev, N.~M.~Bogoliubov}\\
{\it Steklov Institute of Mathematics}
{\it (St.-Petersburg Department)}\\
{\it Fontanka 27, St.-Petersburg, 191023, RUSSIA}
}

\date{}

\maketitle

\def \bal{\boldsymbol\al}
\def \bbe{\boldsymbol\be}
\def \bchi{\boldsymbol\chi}
\def \bdl{\boldsymbol\dl}
\def \Bdl{\boldsymbol\Dl}
\def \bphi{\boldsymbol\phi}
\def \bPsi{\boldsymbol\Psi}
\def \bsi{\boldsymbol\si}
\def \bta{\boldsymbol\eta}
\def \bvphi{\boldsymbol\varphi}
\def \bxi{\boldsymbol\xi}

\def \al{\alpha}
\def \be{\beta}
\def \ga{\gamma}
\def \dl{\delta}
\def \ep{\varepsilon}
\def \nb{\nabla}
\def \tet{\theta}
\def \ka{\varkappa}
\def \la{\lambda}
\def \si{\sigma}
\def \ph{\varphi}
\def \om{\omega}
\def \vt{\vartheta}
\def \z{\zeta}
\def \ta{\theta}

\def \Ga{\Gamma}
\def \Dl{\Delta}
\def \La{\Lambda}
\def \Si{\Sigma}
\def \Ph{\Phi}
\def \Om{\Omega}

\def \cA{\cal A}
\def \cB{\cal B}
\def \cC{\cal C}
\def \cD{\cal D}
\def \cE{\cal E}
\def \cG{\cal G}
\def \cN{\cal N}
\def \cI{\cal I}
\def \cT{\cal T}
\def \cR{\cal R}
\def \cF{\cal F}
\def \cY{\cal Y}
\def \cK{\cal K}
\def \cJ{\cal J}
\def \cX{\cal X}
\def \cZ{\cal Z}
\def \cQ{\cal Q}
\def \cW{\cal W}
\def \cM{{\cal M}}
\def \cL{{\cal L}}
\def \CU{{\cal U}}
\def \cS{{\cal S}}
\def \CK{\mathcal M}
\def\CM{\mathcal M}
\def \CN{\mathcal N}
\def \CV{\mathcal V}

\def \bA{\bold A}
\def \bB{\bold B}
\def \bC{\bold C}
\def \bD{\bold D}
\def \bE{\bold E}
\def \bP{\bold P}
\def \bR{\bold R}
\def \bU{\bold U}
\def \bV{\bold V}
\def \bH{\bold H}

\def\Ba{\rm{B}}
\def\Fa{\rm{F}}
\def \BI{\mathbb{I}}
\def \BC{\mathbb{C}}
\def \BD{\mathbb{D}}
\def \BZ{\mathbb{Z}}
\def \BR{\mathbb{R}}
\def \BQ{\mathbb{Q}}
\def \BN{\mathbb{N}}
\def \IM{\Im}
\def \RE{\Re}
\def \1{^{-1}}
\def \cd{\partial}
\def \at{{\rm arctan}\,}
\def \ch{{\rm ch}\,}
\def \sh{{\rm sh}\,}
\def \th{{\rm th}\,}
\def \bg{{\rm bg}\,}
\def \e{{\rm e}\,}
\def \c{{\rm c}\,}
\def \m{{\rm m}\,}
\def \dr{{\rm d}\,}
\def \o{{\rm o}\,}
\def \ld{\ldots}
\def \bim{{\textbf{\textit M}}}

\def \w{\widetilde}
\def \h{\widehat}
\def \t{\times}
\def \l{\langle}
\def \r{\rangle}
\def \Tr{{\rm Tr}\,}
\def \tr{{\rm tr}\,}
\def \diag{{\rm diag}\,}
\def \row{{\rm row}\,}
\def \Det{{\rm det}\,}
\def \Ddet{\text{\rm Det}\,}
\def \d{\dagger}
\def \pprime{\prime \prime}
\def \babe{\bar\beta}
\def \nt{{\widetilde n}}

\vskip1.5cm
\begin{abstract}
\noindent Relations between the mean values of distributions of flipped spins on periodic Heisenberg $XX$ chain and some aspects of enumerative combinatorics are discussed. The Bethe vectors, which are the state-vectors of the model, are considered both as on- and off-shell. It is this approach that makes it possible to represent and to study the correlation functions in the form of non-intersecting nests of lattice walks and related plane partitions. We distinguish between two types of walkers, namely lock step models and
random turns. Of particular interest is the connection of random turns walks and a circulant matrix. The determinantal representation for the norm-trace generating function of plane partitions with fixed height of diagonal parts is obtained as the expectation of the generating exponential
over off-shell $N$-particle Bethe states. The asymptotics of the dynamical mean value of the generating exponential is calculated in the double
scaling limit provided that the evolution parameter is large. It is shown that the amplitudes of the leading asymptotics depend on the number of diagonally constrained plane partitions.
\end{abstract}


\thispagestyle{empty}

\newpage

\section{Introduction}
\label{sec1}

Some sections of enumerative combinatorics have come to play an important role in the theory of integrable models \cite{bax, fad} and especially in studies of correlation functions \cite{KBI2, ml3}. The exactly solvable Heisenberg $XXZ$ model is a prominent system describing the interaction of spins $1/2$. The $XX$ Heisenberg chain arises from
the $XXZ$ Heisenberg chain
in the limit of vanishing spin interactions along third axis, and it also may be considered as a special free fermion case. It appears that the $XX$ model is related to many problems of theoretical physics and mathematics.
For instance, the model provides a connection between the $XX$ chain and the low-energy QCD, including a possibility of a third order phase transition \cite{gross} in the spin chain \cite{tier, zah}, as well as a
basis for studying entanglement entropy \cite{Sugino}.
Various correlation functions of $XX$ model and their asymptotical behaviour are well studied for a long time \cite{col, col1, col2, ml22, m1}.

The integrable combinatorics \cite{bor, bmumn} attracts now  special attention. In particular, random walks \cite{fish, forr2, forlog} find
interesting implementations in terms of the correlation functions of the quantum exactly solvable models \cite{b1, b11, bmumn}. In \cite{b1, b11} it
was shown that the multi-spin correlation functions over the ferromagnetic vacuum are in one-to-one correspondence with the nests of lattice paths  of
the random turns walkers \cite{fish, forr2}, while the correlation functions calculated over the ground state lead to more complicated
structure of the lattice paths \cite{bmnph, b2, b3}.

Enumeration of plane partitions subject to various constraints is a special part of the enumerative combinatorics \cite{stan1, bres, ver, and, kr, gret}
and fixed volume of diagonal parts is the case of particular interest \cite{st, gan}. For instance, the norm-trace generating function of plane
partitions \cite{st} arises in the study of the temporal evolution of the first moment of particles distribution of the phase model \cite{statm}.
Generalization of the norm-trace generating function, which describes the trace statistics of plane partitions \cite{gan}, results from the partition
function of the four vertex model
in the linearly growing external field \cite{bmjpa}.

In the present paper our approach is based on the theory of symmetric functions \cite{macd}. The Bethe vectors, which are the state-vectors of the model, are considered both as on- and off-shell, allowing to establish
connection between the partitions, the lattice walks, and the plane
partitions \cite{kratt, bres}.
We consider the generating exponential operator, the operator defined by
the sum of projectors to the spin  ``down'' states taken with inhomogeneous weights. The representations for the expectations of the generating exponential over on- and off-shell $N$-particle Bethe states are provided. The generating function
of $N$ random turns walkers is expressed in terms of the entries of products of the circulant matrices \cite{dav, cray}. Ramus's identity \cite{ram} enables us to express
the entries of products of the circulant matrices in terms of the lacunary sums of the binomial coefficients. The multiple series generalizations of Ramus's identity lead to identities
related to enumeration of nests of lattice paths of $N$ vicious walkers. These identities enable to obtain the temporal correlation functions of
inconsecutive flipped spins in terms of the nests of non-intersecting
lattice paths with initial/final sites subject to constraints. The
determinantal representation for the norm-trace generating function of plane partitions
with fixed height of diagonal parts is obtained as the expectation of the
generating exponential over off-shell N-particle Bethe states. The behaviour of the dynamical mean value of the generating exponential is
calculated in the double scaling limit provided that the evolution
parameter is large.

\textit{Organization of the paper}. The outline is given by Section~\ref{sec2}.
The Bethe state-vectors expressed through the Schur polynomials, which admit an interpretation in terms of non-intersecting lattice paths, are presented in Section~\ref{sec5}. Section~\ref{50} is devoted to the transition amplitude over $N$-particle states which is obtained in terms of the circulant matrices. The entries of products of the circulant matrices are obtained in terms of the lacunary sums of the binomial coefficients. The multiple sums generalizations of Ramus's identity are derived in Section~\ref{50}. The norm-trace generating function of the boxed plane partitions with fixed sums of their diagonal parts is derived in Section~\ref{sec53}.
The matrix elements of the Boltzmann weighted generating exponential and their relationship with the non-intersecting lattice walks are considered in
Section~\ref{ss342}. The $N$-particle mean values are estimated in Section~\ref{sec60} for the Boltzmann-weighted generating exponential, for inconsecutive flipped spins, and for powers of the first moment of
the distribution of flipped spins. The estimates at large temporal parameter are obtained which lead to the generating functions of the boxed plane partitions with constrained diagonal parts.
The temporal correlation function of flipped spins is considered in Section~\ref{sec6011} along with its combinatorial interpretation in terms of non-intersecting closed lattice trajectories. Discussion in Section~\ref{sec6} completes the paper.

\section{Outline of the problem}
\label{sec2}

Let us consider the quantum system of interacting spin-$\frac12$ chain consisting of $M$ sites.
Spin ``up'', $\mid \uparrow \rangle_n$, and spin ``down'',
$\mid \downarrow \rangle_n$, states are defined
on $n^{\rm{th}}$ site,
$n\in \{1, 2, \dots, M\}$. Let two operators, ${\sf q}_n$ and ${\bar{\sf q}}_n$,
be the local projectors
(i.e., the spin ``up'' and ``down'' on-site densities) which respect
\begin{equation}
\label{cor:lin413}
{\sf q}_n \mid \downarrow \rangle_n = \mid \downarrow \rangle_n\,,\quad
{\sf q}_n \mid \uparrow \rangle_n = 0\,,\qquad {\bar{\sf q}}_n
\mid \uparrow \rangle_n = \mid \uparrow \rangle_n\,,\quad
{\bar{\sf q}}_n
\mid \downarrow \rangle_n = 0\,.
\end{equation}
Let us introduce the sum of the projectors ${\sf q}_n$ taken with inhomogeneous ``weights'' $\al_n$:
\begin{equation}
\cQ\equiv \sum_{n=1}^M
\al_n {\sf q}_n\,.
\label{cor:lin05}
\end{equation}
The state $\mid\Uparrow\rangle
\equiv \bigotimes_{n=1}^{M} \mid \uparrow \rangle_n$ (spins ``up'' on all sites) is chosen as the reference state (i.e., pseudovacuum \cite{KBI2}),
and therefore the reversed spin on $n^{\rm{th}}$ site $\mid \downarrow \rangle_n$ will be called \textit{flipped} spin. Regarding (\ref{cor:lin413}), the sum $Q(m)\equiv \sum_{k=1}^{m}
{{\sf q}}_k$ is the operator of number of flipped spins on first $m$ sites, and $\cal N \equiv Q(M)$ is the operator of total number of flipped spins.

The paper is devoted to the mean value of the \textit{generating exponential} operator $e^{\cQ}$:
\begin{equation} \prec e^{\cQ}\succ \,
\equiv\, {\sf trace}\,(e^{\cQ}\boldsymbol{\rho} )\,,\qquad \boldsymbol{\rho} \equiv \frac{e^{-\be H}}{{\sf trace}\,(e^{-\be H})}
\,,
\label{cor:lin5}
\end{equation}
where $\be$ is a real positive parameter, $H$ is the Hamiltonian
of the system,
and $\boldsymbol{\rho}$ is the density matrix. The parameter $\be$ might be treated either as an ``evolution'' parameter \cite{b1, statm} or inverse absolute temperature. The symbol ${\sf trace}$ in (\ref{cor:lin5}) implies either summation over $N$-particle (Bethe) states of the model,
${\sf trace} \equiv \tr_N (\cdot)$, or includes additionally summation over $N$, ${\sf trace} \equiv \sum_N \tr_N (\cdot)$, and will be concretized in what follows.

Multiple differentiation of $\prec e^{\cQ}\succ$ with respect to the ``weights'' leads to the correlation
functions of flipped spins, which demonstrate certain combinatorial implications provided that the Bethe state-vectors are expressed in terms of symmetric functions \cite{bmumn}. The approach developed in the present paper is to demonstrate the combinatorial implications of the averages generated by $\prec e^{\cQ} \succ$ (\ref{cor:lin5}).

Generating functions are helpful for derivation of certain correlation functions of the quantum integrable models \cite{KBI2}. The operator $e^{\cQ}$ is called `generating exponential' since
$\prec e^{\cQ}\succ$ (\ref{cor:lin5}) parameterized by the elements of $M$-tuple
${\bf a}_M \equiv (\al_{1},
\al_{2}, \ldots ,
\al_{{M}})$ can be viewed as the
generating function $G ({\bf a}_M) \equiv \prec \exp\cQ ({\bf a}_M)\succ$ of mean values of product ${\varPi}_{\bf k} \equiv \prod_{j=1}^{l} {{\sf q}}_{k_j}$ of flipped spin operators ${\sf q}_n$, not neccisarily at consecutive sites:
\begin{align}
\prec  {\varPi}_{\bf k}\succ
&=\,{\sf trace}\, \bigl({\varPi}_{\bf k}\, \boldsymbol{\rho}\bigr)
\label{cor:lin611}\\
&=\, \lim\limits_{{\textbf a}_{_M} \to 0}
\frac{{\cd}^l\, G ({\bf a}_M)}{\cd \al_{k_1} \cd \al_{k_2}\,\dots\,\cd \al_{k_l}}
\equiv \lim\limits_{{\textbf a}_{_M} \to 0} {\cd}^l_{\al_{k_1} \al_{k_2} \ldots \al_{k_l}} G ({\bf a}_M)\,,
\label{cor:lin61}
\end{align}
where $M \ge k_1 > k_2 > \dots > k_l \ge 1$. Recall that the combinatorial properties of the correlation functions of string $\bar{\varPi}_l \equiv \prod_{j=1}^{l} {\bar{\sf q}}_j$
of the projectors ${\bar{\sf q}}_j$ (\ref{cor:lin413}) have been studied in \cite{bmumn, nest, b3}.

Consider the situation when the elements of ${\bf a}_M$ depend linearly on the site coordinates,
${\bf a}_M = \frac{\al}N \times ({1}, {2}, \ldots , {{M}})$, $\al\in\BR$. The operator ${\cal Q}$ is reduced to ${\cal Q} = \al{\sf M}$, where
\begin{equation}
\label{new30}
{\sf M} \equiv \frac{1}{N}
\sum_{n=1}^{M} n\,{\sf q}_n
\,,\qquad \prec {\sf M}^l \succ \,=\,{\sf trace}\,\bigl({\sf M}^l  \boldsymbol{\rho}\bigr)\,=\,
\mathcal D^l_{\al} \prec e^{\al{\sf M}} \succ\,,
\end{equation}
and $\mathcal D^l_{\al}$ denotes differentiation of $l^{\rm{th}}$ order with respect to $\al$ at $\al=0$. The operator ${\sf M}$ would be considered as the operator of the first moment of the distribution $\frac{n_i}N $ of $N$ flipped spins ($n_i$ is an eigen-value of ${\sf q}_i$). Therefore, the mean value of $e^{\al{\sf M}}$ is the generating function of the mean values of powers of ${\sf M}$.

The temporal evolution of $e^{\al{\sf M}}$ has been studied in \cite{statm} for the quantum phase model,
where ${\sf M}$ is given by (\ref{new30}), ${\sf q}_n$ is on-site boson number operator, and $n_i$ is the occupation number. The operators $e^{h_n{\sf M}}$, $1\le n\le 2N-1$, are used in \cite{bmjpa} to discuss the four vertex model in
inhomogeneous external field. The operator $\cQ$ (\ref{cor:lin05}) is a generalization of ${\sf M}$, and we shall study (\ref{cor:lin5}) and its combinatorial implications using, as a well-developed case, the Heisenberg $XX$ spin chain. The combinatorial implications to be elaborated below are hopefully of interest from the viewpoint of the models mentioned.

Let us introduce the local spin operators
$\si^\pm_n=\frac12 (\si^x_n \pm i\si^y_n)$ and $\si^z_n$ dependent on the lattice argument
$n\in \{1, 2, \dots, M\}$ and acting on the state space  $\frak{H}_M \equiv ({\BC}^2)^{\otimes M}$. The spin operators satisfy the commutation relations:
\begin{equation}
[\si^+_k, \si^-_l]\,=\,\dl_{k l}\,\si^z_l\,,\qquad
[\si^z_k,\si^\pm_l]\,=\,\pm2 \dl_{k l}\,\si^{\pm}_l\,.
\label{comm}
\end{equation}
The definition of spin ``up'' $\mid \uparrow \rangle_n \equiv \Bigl(\begin{matrix} 1\\ 0\end{matrix}\Bigr)_n$
and spin ``down'' $\mid \downarrow \rangle_n \equiv \Bigl(\begin{matrix} 0\\ 1\end{matrix}\Bigr)_n$ states enables that $\si_n^{\pm}$
act on them as the rising/lowering operators:
\begin{equation}
\label{cor:lin411}
\si_n^{+} \mid \downarrow \rangle_n\,=\,\mid \uparrow \rangle_n\,,
\qquad
\si_n^{-} \mid \uparrow \rangle_n\,=\,\mid \downarrow \rangle_n\,,\qquad
\si_n^{-} \mid \downarrow \rangle_n = \si_n^{+}\mid \uparrow \rangle_n =0\,.
\end{equation}
It follows from (\ref{cor:lin411})
that the definitions
\begin{equation}
\label{cor:lin412}
{\sf q}_n \equiv \si_n^{-} \si_n^{+} = \frac12 (1- \si^z_n)\,,
\qquad {\bar{\sf q}}_n \equiv \si_n^{+} \si_n^{-} = \frac12 (1 + \si^z_n)\,,
\end{equation}
ensure (\ref{cor:lin413}).

The Hamiltonian in (\ref{cor:lin5}) is chosen as
that of the $XX$ Heisenberg model:
\begin{eqnarray}
H=H_{\rm xx} -h S^z\,,\qquad H_{\rm xx}\equiv - \,\frac12 \sum \limits^{M}_{n, m=1} \Dl_{nm}\si^+_{n}\si^-_{m}\,, \label{cor:lin2}
\\
S^z=\frac12\sum \limits_{n=1}^M \si^z_n\,, \label{cor:lin3}
\end{eqnarray}
where $\Dl_{nm}$ are given by
\begin{equation}
\Dl_{nm}\,\equiv\,
\dl_{|n-m|, 1} + \dl_{|n-m|, M-1}\,,
\label{cor:lin4}
\end{equation}
${\dl}_{n, l} (\equiv {\dl}_{n l})$ is the Kronecker symbol, $S^z$ is the third component of total spin, and $h\ge 0$ is homogeneous magnetic field.
The number of sites is $M=0\pmod{2}$, the periodic boundary conditions
$\si^{\pm, z}_{n+M}=\si^{\pm, z}_n$, $\forall n\in \{1, 2, \dots, M\}$, are imposed, and $H$ (\ref{cor:lin2}) commutes with $S^z$.
The entries  (\ref{cor:lin4}) constitute so-called
$M\times M$ \textit{transition matrix}
${\bold\Delta}$, which is a special type of \textit{circulant} matrix \cite{dav, cray}).

Depending on the definition of ${\sf trace}$, $\prec e^{\cQ ({\bf a}_M)} \succ$ implies the mean value denoted, in what follows, either $G_{N, \be} ({\bf a}_M)\equiv \l\l e^{\cQ ({\bf a}_M)}  \r\r_{N, \be}$ ($N$ flipped spins and $\be$ are fixed) or $G_{\be} ({\bf a}_M)\equiv\l\l e^{\cQ ({\bf a}_M)} \r\r_{\be}$ ($\be$ is fixed). It will be shown that the combinatorial implications of $G_{N, \be} ({\bf a}_M)$ are related to enumeration of nests of closed lattice trajectories with initial/final positions constrained, as well as
with enumeration of boxed plane partitions with constrained diagonal parts. The implications of $G_{\be} ({\bf a}_M)$ are related to enumeration of random turns walks with constrained initial/final positions.

Recall that $G_\be ({\bf a}_M)$ coincides with the generating function $G(\al, m)$ of the correlation functions of $z$-components of spins for the Heisenberg chains provided that ${\cal Q}=\al Q(m)$
under the choice $\al_1 = \al_2 =\dots = \al_m = \al$ and
$\al_{m+1} = \al_{m+2} =\dots = \al_M = 0$ (so-called, conventional choice) \cite{vk1, kit1, col1, ml3, ess}. The function $G(\al, m)$ has been derived in \cite{corbos} for strongly correlated bosons when $Q(m)$ is the number of particles on a segment of ``length'' $m$.

\section{The state-vectors, the
Schur polynomials and non-intersecting lattice walks}
\label{sec5}

\subsection{The Bethe state-vectors}

Let us introduce a
\textit{strict partition} ${\bmu}=(\mu_1, \mu_2,\,\ldots\,, \mu_N)$ consisting of elements $\mu_k$, $1\leq k\leq N$, called \textit{parts} of ${\bmu}$,
which respect
\begin{equation}
\label{strict2}
M\geq \mu_1> \mu_2 > \, \dots\,> \mu_N \geq 1\,.
\end{equation}
We also introduce the ``staircase'' partition
\begin{equation}
{\bdl}_N \equiv (N, N-1, \dots, 2, 1)\,.
\label{strict}
\end{equation}
The definition (\ref{strict})
allows one to define the partition ${\blad}=(\la_1, \la_2, \dots, \la_N)$ consisting of weakly decreasing non-negative integers:
\begin{equation}
\CK \geq \la_1\geq \la_2\geq \dots\geq
\la_N\geq 0\,, \qquad {\CK}\equiv M-N\,.
\label{strict1}
\end{equation}
The relationship between the parts of $\blad$ and $\bmu$ is expressed as
\begin{equation}
\la_j=\mu_j+j-N-1\,, \qquad 1\le j\le N\,,
\label{eqnpart}
\end{equation}
or $\blad =\bmu - {\bdl}_N$,
where ${\bdl}_N$ is (\ref{strict}). The \textit{volume} of partition, for instance, $\blad$ is the sum of its parts: $|\blad| \equiv \sum_{i=1}^{N} {\la_i}$. The volumes of $\bmu$, $\blad$, and $\bdl$ are related: $|\bmu|=|\blad| + \frac{N}{2} (N+1)$.

Consider the spin chain characterized by $N$ flipped spins on the sites labelled by parts of ${\bmu}$. We define an arbitrary state $|\bmu \rangle$ corresponding to $N$
flipped spins and its conjugate $\langle \bnu |$:
\begin{equation}\label{conwf01}
|\bmu \rangle \equiv \begin{pmatrix}
\prod\limits_{k=1}^N \si_{\mu_k}^{-}\end{pmatrix} \mid
\Uparrow \rangle\,,\qquad\quad
\langle \bnu |\, \equiv \langle \Uparrow \mid \begin{pmatrix}
\prod\limits_{k=1}^N \si_{\nu_k}^{+}\end{pmatrix}\,,
\end{equation}
where $\mid\Uparrow\rangle
\equiv \bigotimes_{n=1}^{M} \mid \uparrow \rangle_n$, and $\si_n^{\pm}$ act on $\mid \uparrow \rangle_n$
and $\mid \downarrow \rangle_n$
according to (\ref{cor:lin411}). The states (\ref{conwf01}) provide a complete orthogonal base:
\begin{equation}
\langle \bnu | \bmu \rangle =
\dl_{\bnu \bmu} \equiv \prod\limits_{n=1}^N
\dl_{\nu_n \mu_n}\,.
\label{tuc3}
\end{equation}

We consider $N$-particle state-vectors as the linear combinations of $| \bmu \rangle$ (\ref{conwf01}), \cite{bmnph, bmumn}:
\begin{equation}
\mid\!\Psi({\textbf u}_N)\rangle \equiv \sum \limits_{\blad \subseteq \{{\CK}^N\}}
S_\blad ({\textbf u}^{2}_N)\,
|\blad + \bdl_N \rangle\,, \label{conwf1}
\end{equation}
where summation is over parts of $\blad =\bmu - {\bdl}_N$
respecting (\ref{strict1}).
The coefficients in (\ref{conwf1}) are given by the {\it Schur polynomials}
$S_\blad$ defined by
the relation \cite{stan1}:
\begin{equation}
S_{\blad} ({\textbf x}_N)\,\equiv\,
\displaystyle{ S_{\blad} (x_1, x_2, \dots , x_N)\,\equiv\, \frac{\det(x_j^{\la_k+N-k})_{1\leq
j, k \leq N}}{\CV({\textbf x}_N)}}\,,
\label{sch}
\end{equation}
where $\CV ({\textbf x}_N)$ is the Vandermonde determinant
\begin{equation} \CV ({\textbf x}_N) \equiv
\det(x_j^{N-k})_{1\leq j, k\leq N}\,=\,
\prod_{1 \leq m< l \leq N}(x_l-x_m)\,.
\label{spxx1}
\end{equation}
The bold notations are used
in (\ref{conwf1}) (and hereafter) for $N$-tuples ${\bf u}^2\equiv (u^2_1, u^2_2, \dots , u^2_N)$, etc. (or ${\bf u}_N^2$, to point out the number of elements).

The state-vectors conjugate of (\ref {conwf01}) are given by
\begin{equation}\label{conj}
\langle \Psi({\bf v}_N) |\,=\,\sum\limits_{\blad \subseteq \{{\CK}^N\}} \langle \blad + \bdl_N |\,
S_\blad ({\textbf v}^{-2}_N)\,.
\end{equation}
The scalar product of the states (\ref {conwf1}) and (\ref {conj}) takes the form:
\begin{equation}
\langle \Psi({\textbf
v}_N)\mid\!\Psi({\textbf u}_N)\rangle\,=\,
\sum_{\blad \subseteq \{{\CK}^N\}}S_\blad
({\textbf v}_N^{-2})S_\blad ({\textbf u}_N^2)\,,
\label{spxx3}
\end{equation}
where the ortho\-go\-na\-li\-ty  (\ref{tuc3}) is used.
Right-hand side of (\ref{spxx3}) is calculated by means of the Cauchy--Binet formula expressed through the Schur polynomials \cite{gant}:
\begin{equation}
\sum_{\blad\subseteq \{L^N\}}S_{\blad}(\mathbf{x}_N) S_{\blad}(\mathbf{y}_N)= \frac{\det T_{L+N}(\textbf{x}_N, \textbf{y}_N)}{
{\CV} ({\textbf x}_N) {\CV} ({\textbf y}_N)}\,, \label{cauchy}
\end{equation}
where summation is over all partitions $\blad$
satisfying: $L\ge \la_1 \ge \la_2\ge \dots \ge \la_N \ge 0$. The matrix $T_{L+N}(\textbf{x}_N, \textbf{y}_N)$ $\equiv$ $(T_{i j}(\textbf{x}_N, \textbf{y}_N))_{1 \le i, j\le N}$ in (\ref{cauchy}) is given by the entries
\begin{equation}
T_{i j}(\textbf{x}_N, \textbf{y}_N)\equiv T_{i j}\equiv {\sf h}_{L+N} (x_i y_j)\,,\qquad {\sf h}_P(x) \equiv \frac{1-x^{P}}{1 - x}\,,
\label{cauchy1}
\end{equation}
where $P\in\BN$. Equations (\ref{cauchy}) and (\ref{cauchy1}) yield the scalar product (\ref{spxx3}):
\begin{equation}
\langle \Psi({\textbf
v}_N)\mid\!\Psi({\textbf u}_N)\rangle\,=\,\frac{1}{{\CV} ({\textbf v}^{-2}_N) {\CV} ({\textbf u}^{2}_N)}\,
\det \Bigl(\frac{1-(u_i/v_j)^{2M}}{1 - (u_i/v_j)^{2}}\Bigr)_{1\le i, j\le N}\,.
\label{spxx}
\end{equation}

Let $\blad$ be a non-strict partition $\la_1\ge\la_2\ge\ldots\ge 0$ of length $l({\blad})=N$. The following identity is valid:
\begin{equation}
\label{ratbe7272}
\sum_{k=1}^N S_{{\blad}+ {\bf e}_{k}} ({\bf x}_N) = \Bigl(\sum_{k=1}^N x_k \Bigr) S_{{\blad}} ({\bf x}_N)\,,
\end{equation}
where ${\bf e}_{k}$, $1\le k\le N$, are $N$-tuples consisting of zeros except of a unity at $k^{\rm th}$ place, say, from left. Generalizations of \eqref{ratbe7272} can be found in  \cite{stan1} (the Murnaghan--Nakayama rule)
and \cite{fult}.

In our case, the parts of $\blad$ are restricted by \eqref{strict1}, since the corresponding parts of $\bmu=\blad+\bdl$
label sites. Let us consider the set $\{S_{{\blad}} ({\bf x}_N)\}_{\blad \subseteq \{{\CK}^N\}}$, whose elements are transformed:
$S_{{\blad}} ({\bf x}_N)
\rightarrow S_{{\blad}\pm {\bf e}_{k}} ({\bf x}_N)$, $k \in \{1, 2, \ldots, N\}$. When several consecutive parts of ${\blad}$ mutually coincide, $S_{{\blad}} ({\bf x}_N)$ is mapped to zero for appropriate $k$ (and sign).
The transformations $S_{{\blad}} ({\bf x}_N)
\rightarrow S_{{\blad}+ {\bf e}_{1}} ({\bf x}_N)$
when $\la_1={\CM}$, or  $S_{{\blad}} ({\bf x}_N)
\rightarrow S_{{\blad}- {\bf e}_{N}} ({\bf x}_N)$
when $\la_N=0$, require a specification.
Consider $N$-tuple ${\bf x}_N\ni x_i$ consisting of $N$ roots ($x_i\ne x_j$ at $i\ne j$) of equation
\begin{equation}
\label{betheexp}
x_i^M=(-1)^{N-1}\,,\qquad 1\le i \le M\,.
\end{equation}
Let $S_{\textrm B}$ to denote $\{S_{{\blad}} ({\bf x}_N)\}_{\blad \subseteq \{{\CK}^N\}}$ with the elements of ${\bf x}_N$ fulfilling
(\ref{betheexp}). Then the transformations are concretized for elements of $S_{\textrm B}$ (due to the definition \eqref{sch}):
\begin{equation}
\begin{array}{c}
S_{(\la_{1}, \la_{2}, \ldots,
\la_{N-1}, 0)} ({\bf x}_N) \rightarrow\, S_{(\la_{1}, \la_{2}, \ldots,
\la_{N-1}, 0)-{\bf e}_{N}} ({\bf x}_N) =
S_{({\CM}, \la_{1}, \la_{2}, \ldots,
\la_{N-1})} ({\bf x}_N)\,,\\
[0.3cm]
S_{({\CM}, \la_{2}, \la_{3}, \ldots,
\la_{N})} ({\bf x}_N) \rightarrow\, S_{({\CM}, \la_{2}, \la_{3}, \ldots,
\la_{N})+{\bf e}_{1}} ({\bf x}_N) =
S_{(\la_{2}, \la_{3}, \ldots,
\la_{N}, 0)} ({\bf x}_N)\,.
\end{array}
\label{betheexp2}
\end{equation}

Equations \eqref{betheexp} coincide with the Bethe equations for the $XX$ model in the exponential form \cite{col1}, $e^{i M\ta_j}=(-1)^{N-1}$, $1\le j \le N$, provided that
the exponential parametrization
${\bf x}_N = e^{i
\bth_N}$ is adopted, where $e^{{i\bth}_N}$ is $N$-tuple $(e^{i\ta_{1}}, e^{i\ta_{2}}, \ldots , e^{i\ta_{N}})$.
Consider $N$-tuple
$\bth_N \equiv (\ta_{1}, \ta_{2}, \dots , \ta_{N})$ consisting of solutions to \eqref{betheexp}:
\begin{equation}
\theta_j = \frac{2\pi
}{M}
\begin{pmatrix}
\displaystyle{
I_j-\frac{N+1}{2}} \end{pmatrix}\,,
\quad 1\le j \le N\,, \label{besol}
\end{equation}
where $M \geq I_1>I_2> \dots>I_N\geq 1$.
The numbers $I_j$ are integer or half-integer depending on whether
$N$ is odd or even, and they constitute $N$-tuple ${{\bf{I}}}_N = (I_1, I_2, \dots, I_N)$. Let us adopt the notation $S_{\rm B}(\bth_N)$ for the set $S_{\rm B}$ to point out
that specific $\bth_N$ is chosen. Thus we come to

\vskip0.3cm \noindent
\noindent{\bf Definition~1:\,}
\textit{The state-vector $|\,\Psi (e^{i{\bth }_N/2})\rangle$ \eqref{conwf1} and its conjugate \eqref{conj} are called $N$-particle Bethe state-vectors of the $XX$ model provided that
the coefficients $S_{{\blad}} (e^{{i\bth}_N}) \in
S_{{\rm B}}(\bth_N)$,
and $|\bmu\r$, $\l \bnu|$ \eqref{conwf01} are given by
$\si^{\pm}_n$ subjected to the periodicity
$\si^{\pm}_{n+M}=\si^{\pm}_n$, $\forall n\in \{1, 2, \dots, M\}$
}.

\vskip0.3cm \noindent
\noindent{\bf Comment:\,}
The variables of the Schur polynomials are chosen in right-hand side of \eqref{conwf1}
as $N$-tuple ${\bf u}^2_N$ just to match to the parametrization of the Bethe equations of more general $XXZ$ Heisenberg chain \cite{bmumn}.
The state-vectors \eqref{conwf1}
parameterised by arbitrary
${\bf u}^2_N$ provide a particular case of so-called \textit{off-shell} Bethe states, in terminology of \cite{slavlec}.

\vskip0.3cm
Let us go over to

\vskip0.3cm \noindent
\noindent{\bf Proposition~1:\,}
\textit{The Bethe state-vectors of $XX$ model $|\,\Psi (e^{i{\bth }_N/2})\rangle$ given by {\sf Definition~1}} $($\textit{on-shell Bethe states} \cite{slavlec}$)$ \textit{are
the eigen-states of $H$ \eqref {cor:lin2} and $S^z$ \eqref {cor:lin3} on periodic chain:
\begin{align}
\label{egv}
\bigr(H_{\rm xx}- h S^z \bigl)\, \mid\!\Psi (e^{i{\bth }_N/2})\rangle\,=\, E_N({\bth }_N)
\mid\!\Psi (e^{i{\bth }_N/2})\rangle\,,\\
\label{egv1}
S^z \mid\!\Psi (e^{i{\bth }_N/2})\rangle\,=\,\Bigr( \frac{M}{2}-N\Bigl)
\mid\!\Psi (e^{i{\bth }_N/2})\rangle\,,
\end{align}
where
\begin{equation}
E_N(\bth)\,=\,-\frac{hM}2 + \sum_{j=1}^N
\ep (\ta_j)\,,\qquad
\ep (\ta_j) \equiv h-
\cos\ta_j\,.
\label{egen}
\end{equation}}

\vskip0.3cm \noindent
\noindent{\bf Proof:\,} Consider the ``exchange'' operator ${\sf{H}}$ introduced by
$H_{\rm xx} \equiv \frac{-{\sf{H}}}2$ \eqref{cor:lin2} and let us obtain its
matrix elements ${\sf{H}} [{\bnu}, {\bmu}] \equiv \l {\bnu} |{\sf{H}} |{\bmu} \r$ between the states \eqref{conwf01}:
\begin{equation}
\label{ratbe727270}
\l {\bnu} |{\sf{H}} |{\bmu} \r = \sum_{k=1}^N \l {\bnu} |{\bmu}- {\bf e}_{k} \r + \l {\bnu} |{\bmu}+ {\bf e}_{k} \r\,,
\end{equation}
where
\begin{align}
\nonumber
&\l {\bnu} |(\mu_1, \mu_2, \ldots, \mu_{N-1}, 1)- {\bf e}_{N} \r = \l {\bnu} |(M, \mu_1, \mu_2, \ldots, \mu_{N-1})\r\,,\\[0.3cm]
\nonumber
&\l {\bnu} |(M, \mu_2, \ldots, \mu_{N-1}, \mu_{N})+ {\bf e}_{1} \r = \l {\bnu} |(\mu_2, \mu_3, \ldots, \mu_{N-1}, \mu_{N}, 1)\r\,.
\end{align}
With regard at \eqref{ratbe7272} and \eqref{betheexp2}, we arrive to the ``matrix eigen-value problem'':
\begin{align}
\label{ratbe727271}
\sum_{\{{\bmu}\}}{\sf{H}} \bigl[{\bnu}, {\bmu}\bigr] S_{{\bmu}-{\dl}} ({\bf x}_N) =
\sum_{k=1}^N S_{{\bnu}+ {\bf e}_{k}-{\dl}} ({\bf x}_N) +
S_{{\bnu} - {\bf e}_{k}-{\dl}} ({\bf x}_N) \\
\label{ratbe727272}
= S_{{\bnu}-{\dl}} ({\bf x}_N) \sum_{k=1}^N \bigl( x_k + x_k^{-1}\bigr) \,,
\end{align}
where $\sum_{\{{\bmu}\}}$ is over
${\bmu}$ \eqref{strict2}. Equality
\eqref{ratbe727271} is due to the orthogonality \eqref{tuc3}, while \eqref{ratbe727272} is due to \eqref{ratbe7272}.
Taking into account
\begin{equation}
\label{ratbe727273}
S_{{\bnu}-{\dl}} ({\bf x}_N) = \l {\bnu} | \Bigl(\sum_{\{{\bmu}\}} S_{{\bmu}-{\dl}} ({\bf x}_N)
|{\bmu} \r\Bigr)\,,
\end{equation}
one straightforwardly obtains that \eqref{egv} and \eqref{egv1} are valid due to \eqref{ratbe727271} and
\eqref{ratbe727272}. $\Box$

\vskip0.3cm
\noindent
The \textit{ground state} solution is given by (\ref{besol}) with ${{\bf{I}}}_N$
substituted by $\bdl_N$ (\ref{strict}):
\begin{equation}\label{grstxx}
\theta^{\,\rm g}_j \equiv \frac{2\pi
}{M}\begin{pmatrix} \displaystyle{\frac{N+1}{2} -j} \end{pmatrix}\,, \quad
1\le j \le N\,.
\end{equation}
Useful relations result from
(\ref{conwf1}), (\ref{conj}),  (\ref{egv}):
\begin{equation}
\begin{aligned}
\langle {\bmu} | e^{-\be H} |\Psi(e^{i{\bth}_N/2})
\rangle & =\, e^{-\be E_N ({\bth}_N) }\,
S_{\blad} (e^{i{\bth}_N}), \\[0.3cm]
\langle \Psi(e^{-i{\bth}_N/2}) | e^{-\be H} |
{\bmu} \rangle &=\, e^{-\be E_N ({\bth}_N) }\,
S_{\blad} (e^{-i{\bth}_N})\,.
\label{ratbe91}
\end{aligned}
\end{equation}

Let us introduce $\CN^2({\textbf u}_N) \equiv \langle \Psi({\textbf
u}_N)\mid\!\Psi({\textbf u}_N)\rangle$ for the scalar product (\ref{spxx}) of the states (\ref{conwf1}) at ${\textbf v}_N={\textbf u}_N$. Then the square of the norm
$\CN^2(e^{i{\bth}_N/2})$ parameterized by solution to the Bethe equations (\ref{betheexp})
takes the following form because of (\ref{cauchy}) and (\ref{spxx}), \cite{bmumn}:
\begin{equation}
\CN^2 (e^{i{\bth}_N/2})\,=\,
\displaystyle{
\frac{M^N}{| \CV (e^{i{\bth}_N}) |^2}
=\frac{M^N}{\prod \limits_{1\leq m<l\leq
N} 2 (1-\cos \frac{2
\pi}{M}(I_l-I_m))}}\,. \label{normxx}
\end{equation}
Decomposition of unity is defined:
\begin{equation}
{\BI}\,=\,\sum\limits_{\{{\bth}_N\}}
\CN^{-2} (e^{i{\bth}_N/2})
\mid\!\Psi (e^{i{\bth}_N/2})\rangle \langle
\Psi (e^{i{\bth}_N/2})\!\mid\,, \label{field7}
\end{equation}
where (\ref {conwf1}) and (\ref {conj}) are taken into account, $\CN^{2} (e^{i{\bth}_N/2})$ is given by (\ref{normxx}), and
summation is over all independent on-shell Bethe states.

\subsection{The Schur polynomials, non-intersecting lattice paths, and plane partitions}

\begin{figure}[h]
\centering
\includegraphics
{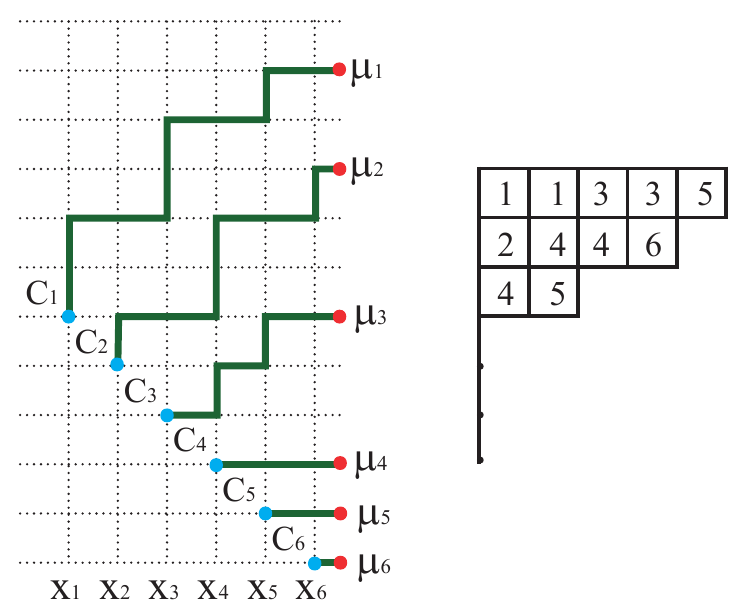}
\caption{A star $\cal{C}$ of $N=6$ lattice paths and semi-standard tableau ${\sf {T}}$ of shape $\blad=(5, 4, 2, 0, 0, 0)$.}
\label{fig:f5}
\end{figure}

The Schur polynomials $S_{\blad}(\mathbf{x}_N)$ (\ref{sch}) admit a combinatorial interpretation since they are  related to the semi-standard Young tableaux, \cite{macd}, which are in one-to-one correspondence with the \textit{nests of non-intersecting lattice paths} (see also Lindstr\"om--Gessel--Viennot lemma, \cite{stan1}). A semi-standard Young tableau ${\sf {T}}$ of shape ${\blad}$
is a diagram possessing $\lambda_i$ cells in $i^{\rm th}$ row ($i=1, \ldots, N$) such that the cells are filled with positive integers $n\in \BN^+$ weakly increasing along rows and strictly increasing downwards along columns (right-hand side of Figure~\ref{fig:f5}).
Consider a grid of vertical and horizontal dashed lines enumerated as in Figure \ref{fig:f5}. \textit{Star} \cite{guttm, 5} is the nest of paths introduced by means of

\vskip0.3cm \noindent
\noindent{\bf Definition~2:\,}
\textit{Star $\cal{C}$, corresponding to semi-standard Young tableau ${\sf {T}}$ of  shape ${\blad}$, is a nest of $N$ non-intersecting lattice paths} (\textit{left-hand side of Figure}~\ref{fig:f5})
\textit{counted from the top of ${\sf {T}}$ and going from points $C_i=(i, N+1-i)$ to points $(N, \mu_i=\la_i+N+1-i), 1\le i \le N$. An $i^{\rm th}$ path makes $\la_i$ upward steps at vertical lines encoded by the integers in $i^{\rm th}$ row of ${\sf {T}}$}.

\vskip0.3cm
\noindent The number $l_j$ of upward steps along the line labelled by $x_j$ coincides with the number of occurrences of $j$ in ${\sf {T}}$. Then, $S_{\blad} (\mathbf{x}_N)$ (\ref{sch}) corresponding to ${\sf {T}}$ of shape ${\blad}$ takes the form:
\begin{equation}\label{schrepr}
S_{\blad} (\mathbf{x}_N) = \sum_{\{\cal{C}\}} \prod_{j=1}^{N}
x_{j}^{l_j},
\end{equation}
where summation is over all admissible stars $\cal{C}$. Let us see how (\ref{schrepr}) enables to obtain (\ref{ratbe7272}). The set of all
semi-standard Young tableau of shapes
${{\blad} \pm {\bf e}_{k}}$, $1\le k\le N$, is characterized by the volume $|{\blad}|\pm 1$. Since $\sum_{i=1}^{N} l_i = |{\blad}|$, one concludes that (\ref{schrepr}) obeys (\ref{ratbe7272}).
The representation (\ref{schrepr}) naturally arises in quantum models soluble by the Quantum Inverse Scattering Method \cite{KBI2}.
The value $S_{\blad} ({\bf 1}_N) \equiv S_{\blad} (1, 1, \ldots, 1)$ gives the number of nests of non-intersecting lattice paths, and it is equal to
\begin{equation}\label{numbpaths1}
S_{\blad} ({\bf 1}_N) = \prod_{1\leq j<k\leq N}\frac{\lambda_j - j - \lambda_k+k}{k-j} = \prod_{1\leq j<k\leq N} \frac{\mu_j - \mu_k}{k-j}\,.
\end{equation}

\begin{figure}[h]
\centering
\includegraphics
{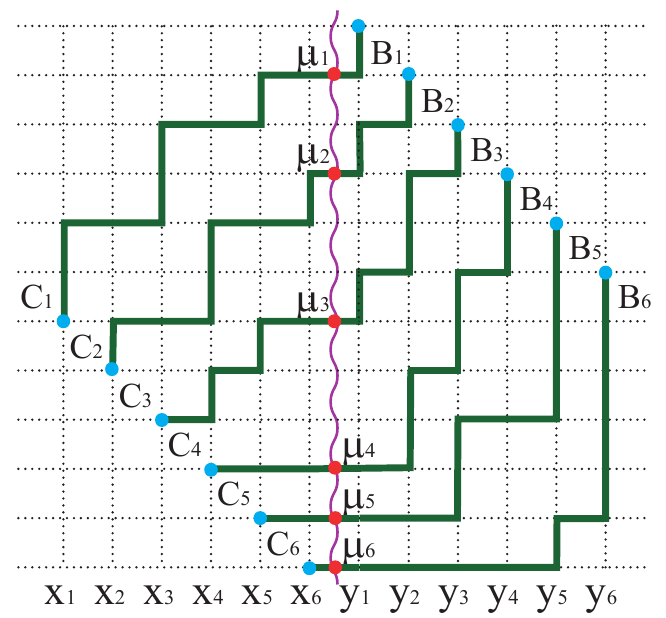}
\caption{\textit{Watermelon} as the nest of lattice paths at  ${\cal M}=6$, $N=6$.}
\label{fig:f6}
\end{figure}

Let us consider the nest of $N$ non-intersecting lattice paths with equidistantly arranged start and end points $C_l$ and $B_{l}$, respectively ($1\le l \le N$). Only upward and rightward steps are allowed for the path in the nest so that an $l^{\rm th}$ one is contained within the rectangle whose lower left and upper right vertices are $C_l$ and $B_l$, respectively. Moreover, the total number ${\CM}=M-N$ of upward steps and the total number $N$ of rightward steps are the same for each path in the nest. The nest described is called \textit{watermelon}
(cf. Figure~\ref{fig:f6}), \cite{guttm}.

As far as watermelon configurations are concerned, a \textit{conjugate} nest ${\cal B}$ of lattice paths for the Young tableau of shape $({\CM}-\la_1, {\CM}-\la_2, \ldots, {\CM}-\la_N)$ has been introduced in \cite{bmumn}.
A typical watermelon in Figure~\ref{fig:f6} can be viewed (with a reference to \cite{bmumn} for more details) as a star ${\cal C}$ ``glued'' to conjugate star ${\cal B}$ along the points on the dissection line (wavy line in Figure~\ref{fig:f6}). The partition $\bmu\ni\mu_l$ \eqref{strict2} determines the ordinates of the points $(N, \mu_l)$, which must coincide, as the end points of the star ${\cal C}$ (cf. Figure~\ref{fig:f5}), with those characterizing the conjugate star ${\cal B}$, \cite{bmumn}. The Schur polynomial corresponding to the conjugate nest is
\begin{equation}\label{rsf}
S_{\blad} ({\bf y}_N) \equiv
S_{\blad} (y_1, y_2, \ldots, y_N) = \sum_{\{\cal{B}\}} \prod_{r=1}^{N}
y_{r}^{{\cal M}-b_r},
\end{equation}
where $b_r$ is the number of upward steps along $y_{r}$, and summation is over all stars $\cal{B}$.

Under the $q$-parametrization
\begin{equation}
\label{rep21}
\textbf{v}^{-2}={\bf q}_N\equiv (q, q^2, \dots, q^N)\,,\qquad
\textbf{u}^2 = {{\bf q}_N}/q
\,,
\end{equation}
the scalar product (\ref{spxx3})
is expressed:
\begin{equation}
\langle \Psi(\textbf{q}^{-1/2}_N)|
\Psi((\textbf{q}_N/{q})^{1/2})
\rangle\,=\,
\sum_{\blad \subseteq \{{\CK}^N\}} S_{\blad} ({\bf q}_N) S_{\blad} \Bigl(\frac{{\bf q}_N
}{q}\Bigr)
\label{spxx03}
\end{equation}
($S_{\blad} ({\bf q}_N/q)$
is so-called `principal specialization' of the Schur polynomial, see \cite{stan1}). Right-hand side of  (\ref{spxx03}) is the generating function of the number of watermelons characterized by the points $C_l$ and $B_{l}$ ($1\le l \le N$):
\begin{equation}
\lim_{q\to 1}\, \langle \Psi(\textbf{q}^{-1/2}_N)|
\Psi((\textbf{q}_N/{q})^{1/2})
\rangle \,=\, \sum_{\bmu \subseteq \{{M}^N\}}
\Bigl(\sum_{\{\cal{C}\}_{\bmu}} 1\Bigr) \Bigl(\sum_{\{\cal{B}\}_{\bmu}} 1\Bigr)\,,
\label{schrepr5}
\end{equation}
where (\ref{schrepr}) and (\ref{rsf}) are used, and the notations $\{\cal{C}\}_{\bmu}$, $\{\cal{B}\}_{\bmu}$ are to stress that the summation is over the nests characterized by specific $\bmu$ \cite{bmumn}.

\begin{figure}[h]
\center
\includegraphics [scale=0.5]{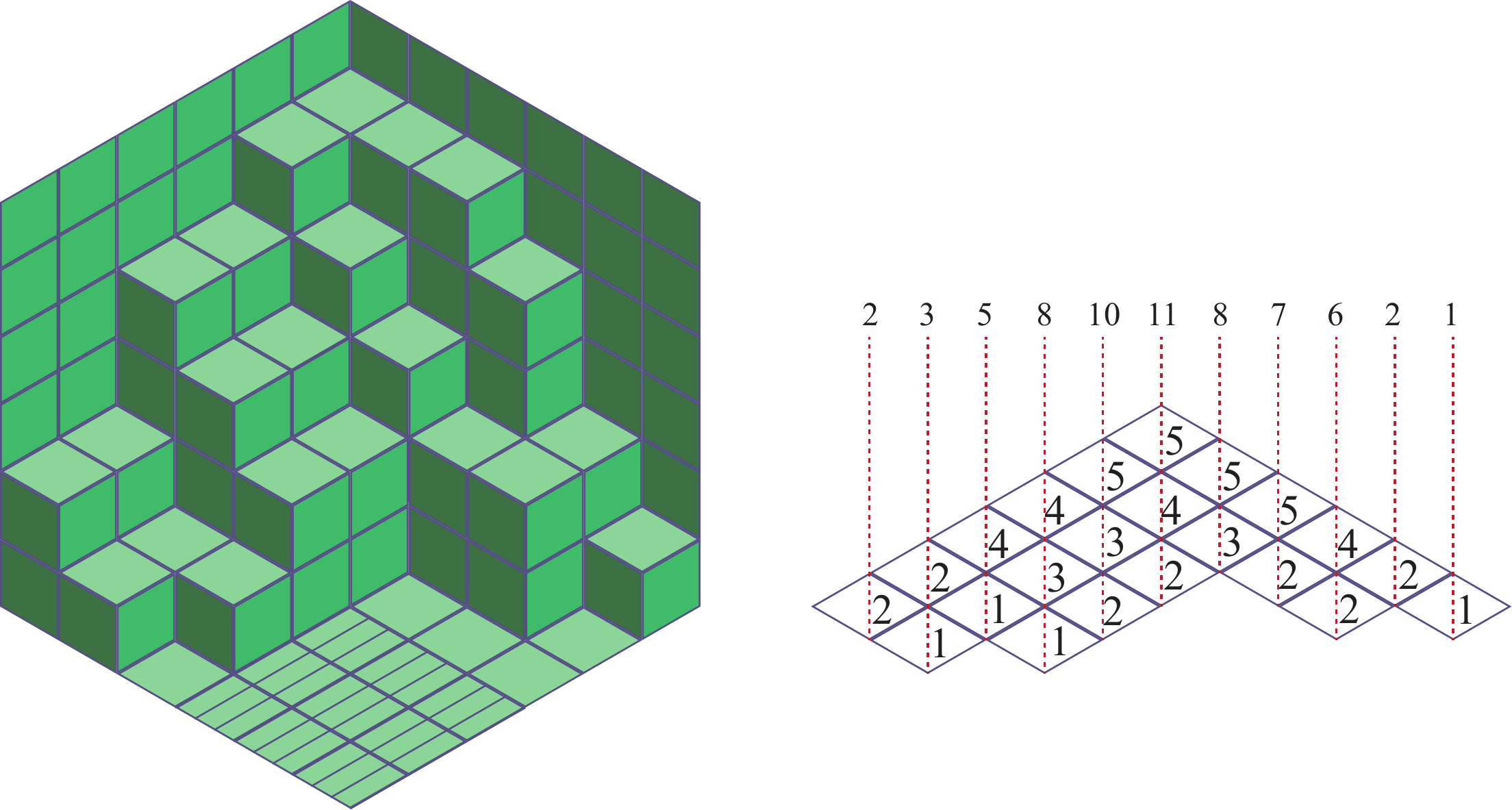}
\caption{Plane partition with $|\blad|=11$ equivalent to watermelon in Figure \ref{fig:f6}.}
\label{fig:ff9}
\end{figure}

A \textit{plane partition} ${\bpi}$ contained in ${N\times N\times {\cal M}}$ box ${\cal B} (N, N, {\cal M})$
is an array $({\pi}_{ij})_{i, j\ge 1}$ of non-negative
integers that satisfy
${\pi_{ij} \leq {\cal M}}$, ${\pi_{ij} \geq \pi_{i+1, j}}$ and ${\pi_{ij} \geq \pi_{i, j+1}}$ for all ${i, j \geq 1}$.
Furthermore, ${\pi_{ij}=0}$ whenever ${i}$ or ${j}$ exceed ${N}$ (Figure~\ref{fig:ff9}), \cite{bres, stan1}. There exists bijection between the watermelon configuration of non-intersecting lattice paths (Figure~\ref{fig:f6}) and the plane partition in ${\cal B} (N, N, {\cal M})$ (Figure~\ref{fig:ff9}).
The trace of $s^{\rm th}$ diagonal of plane partition
counted from lower left corner
is $\tr_s {\bpi}\equiv \sum_{N+j-i=s} \pi_{i j}$, $1\le s\le 2N-1$. The \textit{volume} of ${\bpi}$ is $|{\bpi}| = \sum_{s=1}^{2N-1} {\tr}_s {\bpi}$.  The bijection is such that the heights of diagonal columns are given by parts of
${\blad}$, and thus $\tr_N {\bpi}=|{\blad}|$ (all traces are depicted in Figure~\ref{fig:ff9}; notice that $3\times 3$ square hatched in Figure~\ref{fig:ff9} is commented in Sec.~\ref{sec555}).

\section{The transition amplitude and random turns walks of vicious walkers}
\label{50}

\subsection{Multi-particle transition amplitude}
\label{s341}

\begin{figure}[h]
\centering
\includegraphics[scale=1.0]{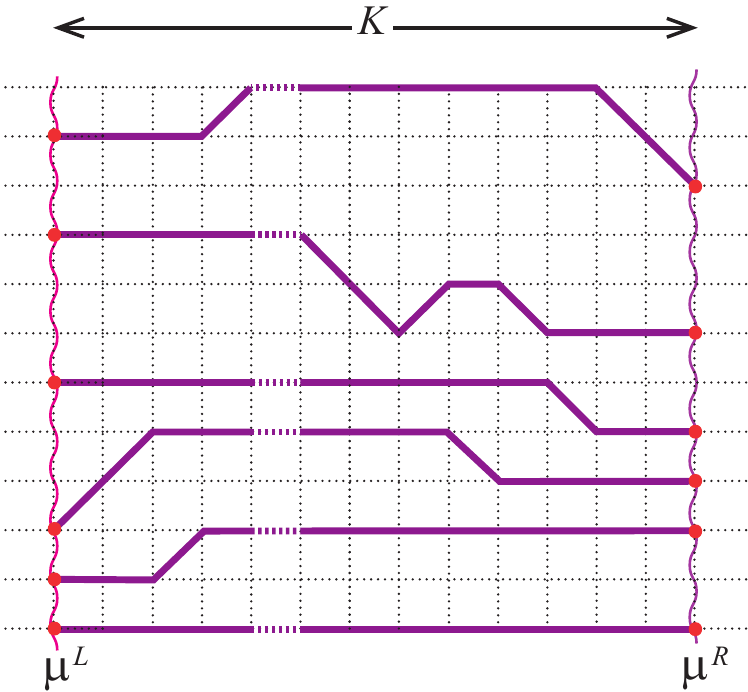}
\caption{Random turns vicious walkers.}
\label{fig:f1}
\end{figure}

One-dimensional random walks of \textit{vicious walkers}, who annihilate one another whenever they meet at the same lattice site, attract attention after \cite{fish}, and so-called {\it lock step}, \cite{forr1}, and {\it random turns} models, \cite{forr2, forr},
are distinguished. Suppose that there are $N$ walkers on a one-dimensional lattice. In the random turns model only a single randomly chosen walker jumps
at each tick of a clock
to either of adjacent sites while the remaining walkers are staying (a typical example in Figure~\ref{fig:f1}, the discrete time in the horizontal direction). It has been proposed in \cite{b1, b11} to use the Heisenberg $XX$ chain
in order to interpret random movements in the random turns model as transitions between spin ``up'' and ``down'' states.

The generating function of the lattice trajectories of $N$ random turns vicious
walkers is given by $N$-particle \textit{transition amplitude}
between the states $\langle {\bmu^L} |$ and $| {\bmu^R} \rangle$ parameterized by parts of ${\bmu^L} \equiv (\mu^L_1, \mu^L_2, \dots, \mu^L_N)$ and ${\bmu^R}\equiv(\mu^R_1, \mu^R_2, \dots, \mu^R_N)$ interpreted as
initial and final positions of the walkers:
\begin{equation}
G_{{\bmu^L}; {\bmu^R}}(\be) \equiv \langle {\bmu^L} |\, e^{- \be H_{\rm xx}+ \be h S^z} |\, {\bmu^R} \rangle\,,
\label{mpcf}
\end{equation}
where the Hamiltonian (\ref{cor:lin2}) is used. The representation
(\ref{mpcf}) is re-expressed:
\begin{align}
G_{{\bmu^L}; {\bmu^R}}(\be) &=\,
e^{\be h (\frac{M}2 - N)}
G^{\,0}_{{\bmu^L}; {\bmu^R}} (\be) \,,
\label{mpcf21} \\[0.2cm]
G^{\,0}_{{\bmu^L}; {\bmu^R}} (\be) & \equiv
\langle {\bmu^L} |\, e^{- \be H_{\rm xx}} |\, {\bmu^R} \rangle\,,
\label{mpcf2}
\end{align}
where the commutation relation
\begin{equation}
\nonumber
e^{\be h S^z} \si^\pm_n\,=\,
e^{\pm\be h} \si^\pm_n e^{\be h S^z}
\end{equation}
is used together with $S^z \mid \Uparrow \rangle = \frac M2 \mid \Uparrow \rangle$. The exponential factor on right-hand side of (\ref{mpcf21}) is due to coupling of the spin chain to homogeneous magnetic field, and the corresponding exponent is proportional to the eigen-value (\ref{egv1}) of the total spin.

Let us obtain equation, which governs the transition amplitude $G_{{\bmu^L}; {\bmu^R}} (\be)$ (\ref{mpcf}). Indeed, differentiating (\ref{mpcf21}) over $\be/2$ and using \eqref{ratbe727270}, \eqref{ratbe727271},
one obtains at fixed ${\bmu^L}$:
\begin{align}
\label{mpcf333}
\frac {d\, G_{{\bmu^L}; {\bmu^R}}(\be)}{d (\be/2)} &=\,
\sum_{\{\bnu\}} G_{{\bmu^L}; {\bnu}} (\be)\,
{\sf H}\bigl[{\bnu}, {\bmu^R}\bigr]
+ h ({M} - 2N)\, G_{{\bmu^L}; {\bmu^R}}(\be) \\
\label{mpcf3}
&=\,\sum_{k=1}^N \bigl(G_{{\bmu^L}; {\bmu^R}+{\bf e}_{k}} (\be)  + \,G_{{\bmu^L}; {\bmu^R}-{\bf e}_{k}} (\be)\bigr)
+ h ({M} - 2N)\, G_{{\bmu^L}; {\bmu^R}}(\be)
\,,
\end{align}
where ${\sf{H}} \bigl[{\bnu}, {\bmu^R}\bigr]$ is defined by \eqref{ratbe727271}
(and a similar equation for fixed ${\bmu^R}$). Equation (\ref{mpcf3}) is supplied with the initial condition $G_{{\bmu^L}; {\bmu^R}}(0) = \dl _{{\bmu^L} {\bmu^R}}$ (see \eqref{tuc3}), as well as with the periodicity condition:
\begin{equation}
\label{ratbe7171}
G_{{\bmu^L}; {\bmu^R}} (\be) = G_{{\bmu^L}+ M {\bf e}_{k};\, {\bmu^R}} (\be) = G_{{\bmu^L};\, {\bmu^R}+ M {\bf e}_{k}} (\be)\,,\qquad \forall k\in \{1, 2, \dots, M\}\,.
\end{equation}
Non-intersecting walks of vicious walkers are described by solution to (\ref{mpcf3}) provided that the non-intersection condition is imposed: $G_{{\bmu^L}; {\bmu^R}}(\be)=0$, if $\mu^R_k=\mu^R_p$ (or $\mu^L_k=\mu^L_p$) for any $k \ne p$, $1\leq k, p\leq N$.

The orthonormality relation is valid for the Schur polynomials (\ref{sch}):
\begin{equation}
\label{ratbe77}
\frac{1}{M^N}
\sum_{\{{\bphi}_N\}} |\CV (e^{i{\bphi}_N})|^2\,
S_{{\blad^L}}(e^{-i{\bphi}_N})\,
S_{{\blad^R}}(e^{i{\bphi}_N})
= \dl_{{\blad^L} {\blad^R}} \,,
\end{equation}
where $\dl_{{\blad^L} {\blad^R}}$ is given by \eqref{tuc3}.
The sum in (\ref{ratbe77}) is over $N$-tuples ${\bphi}_N = (\phi_{k_1}, \phi_{k_2}, \dots, \phi_{k_N})$,
where $\phi_n = \frac{2\pi}{M} \bigl(n-\frac{M}{2}\bigr)$ and $M\ge k_1>k_2> \cdots > k_N\ge 1$. Moreover, ${\CV} (e^{i {\bphi}_N})$ is defined by (\ref{spxx1}), and
$e^{\pm i {\bphi}} \equiv (e^{\pm i\phi_1}, e^{\pm i\phi_2}, \dots, e^{\pm i\phi_N})$. One obtains
from (\ref{ratbe7272}) and (\ref{ratbe77}) (cf. \cite{nest}) the following

\vskip0.3cm \noindent
\noindent{\bf Proposition~2:\,}
\textit{Solution to \eqref{mpcf3} respecting the initial condition $G_{{\bmu^L}; {\bmu^R}}(0) = \dl_{{\blad^L} {\blad^R}}$, as well as the periodicity condition
\eqref{ratbe7171}, is given by}
\begin{equation}
G_{{\bmu^L}; {\bmu^R}} (\be)\,=\,
\displaystyle{\frac{1}{M^N}
\sum\limits_{\{{\bphi}_N\}}
e^{- \be E_N({\bphi}_N)}}|{\CV} (e^{i {\bphi}_N})|^2\,
S_{{\blad^L}}(e^{i {\bphi}_N})
S_{{\blad^R}}(e^{-i {\bphi}_N})\,,
\label{ratbe7}
\end{equation}
\textit{where
${\blad^{L, R}}={\bmu^{L, R}}- {\bdl}_N$, $E_N({\bphi}_N)$ is defined by \eqref{egen}, and the sum is the same in \eqref{ratbe77}. The determinantal representation of $G_{{\bmu^L}; {\bmu^R}} (\be)$ \eqref{ratbe7} ensures the
non-intersection property}:
\begin{equation}
G_{{\bmu^L}; {\bmu^R}} (\be) =
e^{\be h(\frac{M}{2}-N)} G^0_{{\bmu^L}; {\bmu^R}} (\be)\,,\quad
G^0_{{\bmu^L}; {\bmu^R}}(\be) = \displaystyle{
\det \bigl(
{G}^0_{ {\mu_{n}^L}; {\mu_{k}^R}} (\be) \bigr)_{1\le
n, k \le N}}\,,
\label{ratbe6}
\end{equation}
\textit{where ${G}^0_{j;\,m}(\be)$ is the solution to \eqref{mpcf3} at $N=1$ and $h=0$:
\begin{equation}
\label{qanal277}
{G}^0_{j;\,m}(\be) \,=\, \displaystyle{
\frac{1}{M} \sum\limits_{n=1}^{M}
e^{\be \cos \phi_n}\,
e^{i \phi_n(m-j)}}\,.
\end{equation}
}

\vskip0.3cm \noindent
\noindent{\bf Proof:\,}
The representation \eqref{ratbe7} fulfils \eqref{mpcf3} due to \eqref{ratbe727271} and
\eqref{ratbe727272}. Definition of determinant allows to obtain \eqref{ratbe6} from \eqref{ratbe7}, \cite{forr2, b1, nest}. $\Box$

The function ${G}^0_{j;\,m}(\be)$ (\ref{qanal277}) is reduced to the modified Bessel function of the first kind provided that $\frac{1}{M} \sum_{n=1}^{M}$ is approximately replaced by $\frac{1}{2\pi} \int_{-\pi}^{\pi} dp$ at large enough $M\gg 1$:
\begin{equation}
\label{ratbe7373}
{G}^0_{j;\,m}(\be) \,=\,I_{|j- m|}(\be)\,.
\end{equation}
The power series is valid for $I_{|j-m|}({\be})$:
\begin{equation}
I_{|j-m|}({\be})=\sum_{K\geq|m-j|}
\frac{({{\be}}/2)^K}{\bigl( \frac{K-|j-m|}2 \bigr)!\, \bigl(\frac{K+|j-m|}2 \bigr)!}\,,
\label{25}
\end{equation}
where $K+|j-m|= 0 \pmod{2}$.
Let us recall that $\mathcal D^K_{s}$ is the differentiation of $K^{\rm{th}}$ order with respect to $s$ at $s=0$ (cf.~(\ref{new30})). Applying $\mathcal D^K_{\be/2}$ to
${G}^0_{j;\,m}(\be)$ \eqref{ratbe7373} and using \eqref{25}, one obtains the number ${|P^0_K(m\rightarrow j)|}$, \cite{guttm}, of $K$-step paths between $m^{\rm th}$ and $j^{\rm th}$ sites in terms of the binomial coefficient \cite{b1}:
\begin{equation}
|P^0_K(m\rightarrow j)| =
\begin{pmatrix}
2L+|m-j| \\
L
\end{pmatrix}\,,
\label{26}
\end{equation}
where $L$ is one-half of the total number of turns:
$L\equiv(K-|m-j|)/2$.

\subsection{The random turns walks and the circulant matrix}
\label{ss341}

Acting by $\mathcal D^K_{\be/2}$ on $G_{{\bmu^L}; {\bmu^R}} (\be)$ (\ref{mpcf}) one obtains the matrix element for $K^{\rm th}$ power of the total Hamiltonian:
\begin{equation}
\mathfrak{G} ({\bmu^L}; {\bmu^R}\,| K) \equiv \mathcal D^K_{\be/2} \,G_{{\bmu^L}; {\bmu^R}} (\be)
= \langle {\bmu^L} |\,(-2 H)^K  |\, {\bmu^R} \rangle
\label{mpcf44}\,.
\end{equation}
It follows from (\ref{mpcf44}) that
$\mathfrak{G} ({\bmu^L}; {\bmu^R}\,| 0) = \dl_{{\bmu^L} {\bmu^R}}$
due to the orthogonality (\ref{tuc3}). Let us represent the solution to (\ref{mpcf3}) as the power series:
\begin{equation}\label{new1}
G_{{\bmu^L}; {\bmu^R}} (\be)\,=\,
\sum_{K=0}^{\infty}
\frac{(\be/2)^K}{K!}\,
\mathfrak{G} ({\bmu^L}; {\bmu^R}\,| K)\,,
\end{equation}
where the coefficients $\mathfrak{G} ({\bmu^L}; {\bmu^R}\,| K)$ (\ref{mpcf44}) respect the equation:
\begin{align}
\nonumber
\mathfrak{G} ({\bmu^L}; {\bmu^R}\,| K+1) & = h(M-2N)\, \mathfrak{G} ({\bmu^L}; {\bmu^R}\,| K) \\
& +\, \sum_{k=1}^{N} \bigl(\mathfrak{G} ({\bmu^L}; {\bmu^R}+{\bf e}_{k}\,| K) +
\mathfrak{G} ({\bmu^L}; {\bmu^R}-{\bf e}_{k}\,| K)\bigr)\,.
\label{mpcf6}
\end{align}
Equation (\ref{mpcf6}) is supplied with the initial condition $\mathfrak{G} ({\bmu^L}; {\bmu^R}\,| 0) = \dl_{{\bmu^L} {\bmu^R}}$, as well as with appropriate periodicity and non-intersection requirements.

Equation (\ref{mpcf6}) at $h=0$
is an isotropic version of more general equation derived in \cite{forr2} for the random turns model with anisotropy of the shifts $\pm {\bf e}_k$, $1\le k\le N$. A comparison with \cite{forr2} demonstrates that not only jumps to neighboring sites
are allowed, but there is an opportunity, due to the homogeneous magnetic field $h$, for all walkers to stay stationary (cf. Figure~\ref{fig:f1}).

Let us assume that $G^0_{{\bmu^L}; {\bmu^R}} (\be)$ (\ref{mpcf2}) respecting (\ref{mpcf3}) at $h=0$
is also given by the series analogous to (\ref{new1}).
The corresponding coefficients  $\mathfrak{G}^{\,0}({\bmu^L}; {\bmu^R}\,| K)$ defined as follows,
\begin{equation}
\mathfrak{G}^{\,0}({\bmu^L}; {\bmu^R}\,| K)
\equiv \mathcal D^K_{\be/2}\, G^{\,0}_{{\bmu^L}; {\bmu^R}} (\be) = \langle {\bmu^L} |\,(-2 H_{\rm xx})^K  |\, {\bmu^R} \rangle\,,
\label{qanal13}
\end{equation}
respect (\ref{mpcf6}) at $h=0$.
Expanding the exponential in (\ref{mpcf21}) and taking (\ref{new1}) into account, one obtains the identity:
\begin{equation}
\mathfrak{G} ({\bmu^L}; {\bmu^R}\,| K)\,
 =\,\sum_{p=0}^{K}
\begin{pmatrix}
K\\
p
\end{pmatrix}
\bigl(h(M-2N) \bigr)^p\,
\mathfrak{G}^{\,0}({\bmu^L}; {\bmu^R}\,| K-p)\,,
\label{mpcf4}
\end{equation}
where $\begin{pmatrix}
K\\ p
\end{pmatrix}$ is the binomial coefficient. Right-hand side of (\ref{mpcf4}) is reduced at $h=0$ to $\mathfrak{G}^{\,0} ({\bmu^L}; {\bmu^R}\,| K)$ since only $p=0$ contributes.

The circulant matrix
${\bold\Delta}$ (\ref{cor:lin4}) leads to $N=1$ solution of (\ref{mpcf6}) at $h=0$:
\begin{equation}
\mathfrak{G}^0 (j, m | K) =  \langle \Uparrow \mid \si_j^{+} (-2 H_{\rm xx})^K \si_m^{-} \mid
\Uparrow \rangle = \bigl({\bold \Delta}^K \bigr)_{j m}\,,
\label{avv}
\end{equation}
where
$\bigl({\bold \Delta}^K \bigr)_{j m}$ is the entry of $K^{\rm th}$ power of
${\bold\Delta}$, which obeys
\begin{equation}
\label{dmcf}
\bigl({\bold \Delta}^{K+1} \bigr)_{j m}
= \bigl({\bold \Delta}^{K} \bigr)_{j, m+1} +
\bigl({\bold \Delta}^{K} \bigr)_{j, m-1}\,.
\end{equation}
The initial condition is fulfilled since $\mathfrak{G}^0 (j, m | 0)$ is the Kronecker symbol $\delta_{j m}$. The periodicity is also consistent with the circulant matrix (\ref{cor:lin4}).

Position of the walker on the chain is labelled by the spin ``down''
state, while the empty sites correspond to spin ``up'' states. Let $|P^0_K(j \rightarrow m)|$ to denote the number of $K$-step paths of a single walker between $j^{\rm {th}}$ and $m^{\rm {th}}$ sites ($h=0$). Evaluation of (\ref{qanal13}) corresponding to $N=1$ results in $|P^0_K(j \rightarrow m)|= \bigl({\bold\Delta}^K \bigr)_{j m}$ in agreement with (\ref{avv}).

Let us turn to the lattice paths made by $N$ random turns vicious walkers at $h=0$ with initial and final positions arranged as the strict partitions ${\bmu^L}$ and ${\bmu^R}$, respectively. Let
$|P^0_K ({\bmu^L} \rightarrow\, {\bmu^R})|$ be the number of nests of $K$-step paths and assume that a single walker
jumps to adjacent site at each step. We formulate the following

\vskip0.3cm \noindent
\noindent{\bf Proposition~3:\,}
\textit{The number of nests of non-intersecting lattice paths of $N$ random turns vicious walkers with $K$ steps is equal to the amplitude $\mathfrak{G}^{\,0} ({\bmu^L}; {\bmu^R}\,| K)$ solving \eqref{mpcf6} at $h=0$}:
\begin{align}
\nonumber
|P^0_K ({\bmu^L} \rightarrow\,{\bmu^R})| & =\,
\mathfrak{G}^{\,0}({\bmu^L}; {\bmu^R} | K) \\
\label{qanal272}
& =\,\sum_{|{\bf n}|=K} P({\bf n})\,
\det\bigl(({\bold\Delta}^{n_j} )_{\mu^L_{i}; \mu^R_j}\bigr)_{1
\le i, j \le N}\,,
\end{align}
\textit{where ${\bf n}=(n_1, n_2, \ldots, n_N)$, $|{\bf n}|\equiv n_1+n_2+ \ldots + n_N$, $P({\bf n})$ is the multinomial coefficient},
\begin{equation}
\label{qanal271}
P({\bf n}) \equiv
\frac{(n_1+n_2+ \ldots+n_N)!}{n_1!\, n_2!\, \dots n_N!}\,,
\end{equation}
\textit{the entry $({\bold \Delta}^{n})_{j m}$ is defined by \eqref{avv}, and $({\bold \Delta}^{0})_{j m} = {\delta}_{j m}$}.

\vskip0.3cm \noindent
\noindent{\bf Proof:\,} See \textsf{Appendix I}. $\Box$

Equation $\mathfrak{G} ({\bmu^L}; {\bmu^R} | K)$ (\ref{mpcf4}) demonstrates that either a single walker chosen randomly jumps to one of adjacent sites with equal probabilities or all walkers are staying stationary. Right-hand side of (\ref{mpcf4}) is the polynomial of a single variable $h(M-2N)$. The coefficients
$\mathfrak{G}^{\,0}({\bmu^L}; {\bmu^R} | K-p)$ enumerate, due to \textsf{Proposition~3}, $(K-p)$-step nests of paths of $N$ walkers. In turn, the number $\begin{pmatrix}
K\\ p
\end{pmatrix}$ of $p$-element combinations of the set of $K$ steps enumerates all the possibilities for $N$ walkers to stay stationary $p$ times. A typical nest of $N=6$ paths is shown in Figure~\ref{fig:f1} ($K=13$, $p=1$) where dashed lines imply that walkers are staying. As far as $|P_{K-p}^0 ({\bmu^L} \rightarrow\, {\bmu^R})|$ (\ref{qanal272}) is concerned, the nest in Figure~\ref{fig:f1} corresponds to $n_1=0$, $n_2=1$, $n_3=3$, $n_4=1$, $n_5=4$, $n_6=3$.

\subsection{Generalized
Ramus's identity}
\label{s342}

The present section is devoted to a relationship between the powers of the circulant matrix ${\bold\Delta}$ \eqref{cor:lin4} and the binomial coefficients \eqref{26}.

Calculation of the entries of integer positive powers of circulant matrices attracts attention \cite{rim1, rim2, feng}. For instance, the entries
$\bigl({\bold \Delta}^K \bigr)_{j m}$ at $K$ arbitrary are obtained in \cite{rim1, rim2} for ${\bold \Delta}$ of even order in terms of the Chebyshev polynomials. In the present paper it is proposed to express $\bigl({\bold\Delta}^K \bigr)_{j m}$ by means of \textit{Ramus's identity} \cite{ram} (cf. \cite{ram1, ram2}):
\begin{equation}
\label{rams}
\frac{2^{n}}{R}\sum_{j=0}
^{R-1}
\cos^n\frac{\pi j}{R}
\cos\frac{\pi j (n-2t)}{R}
\,=\,\sum_{l=0, 1, 2, \ldots}
\begin{pmatrix} n \\ t + R\cdot l \end{pmatrix}
\,,\quad 0\le t < R \,.
\end{equation}
The latter provides the entries
in question in terms of the binomial coefficients thus stressing the connection with enumeration of the lattice walks.

The vanishing $\bigl({\bold\Delta}^K \bigr)_{j m}=0$ occurs for the circulant matrix (\ref{cor:lin4}) in the case $K-|j-m| = 1 ({\rm mod}\,2)$. In the case $K - |j-m| =0 ({\rm mod}\,2)$, Ramus's identity (\ref{rams}) allows us to formulate

\vskip0.3cm \noindent
\noindent{\bf Proposition~4:\,}
\textit{The row-column indices $j, m$ of $M\times M$ matrix respect $|j-m|\le M-1$. Let us consider $L\equiv \frac{K-|j-m| + p M}{2}$, where $p\in{\mathbb Z}$ is chosen so that $0\le L\le \frac{M}{2}$. Then},
\begin{equation}
\label{qanal276}
\bigl({\bold\Delta}^K \bigr)_{j m}\, =\,
\begin{pmatrix} K \\ L {\bar\dl}_{L, \frac{M}{2}}
\end{pmatrix}_{{M}/{2}}\,,
\end{equation}
\textit{where ${\bar\dl}_{L, \frac{M}{2}}\equiv 1-{\dl}_{L, \frac{M}{2}}$, and
the notation for the lacunary sum of binomial coefficients is used} \cite{ram3}:
\begin{equation}
\begin{pmatrix} K \\ L \end{pmatrix}_{{M}/{2}} \equiv
\sum_{n=0, 1, 2, \ldots}
\begin{pmatrix} K \\ L +\frac{M}{2}\cdot n \end{pmatrix} \,.
\label{qanal27611}
\end{equation}

\vskip0.3cm \noindent
\noindent{\bf Proof:\,}
The transition element (\ref{qanal13}) arising from (\ref{qanal277}) takes the form
(recall that $M$ is even):
\begin{align}
\mathfrak{G}^{\,0}(j; m | K)
= \frac{2^{K+1}}{M}
\sum_{l=0}^{\frac{M}{2}-1}
\cos^K \Bigl(\frac{2\pi l}{M}\Bigr)
\cos\Bigl(\frac{2\pi l |m-j|}{M}\Bigr)
\nonumber \\
+\,\bigl((-1)^{K+|m-j|} - 1\bigr)
\frac{2^{K}}{M}\,.
\label{qanal278}
\end{align}
Ramus's identity \eqref{rams} allows us to re-express the series in (\ref{qanal278}) provided that $n$ and $n-2t$ are replaced by $K$ and $K-2L$, respectively. As the result, the validity of (\ref{qanal276}) is verified for $K - |m-j|=0 ({\rm mod}\,2)$ at $L\ne \frac{M}{2}$
(cf. \textsf{Appendix II} for illustrative examples). It is clear from (\ref{rams}) and (\ref{qanal278}) that (\ref{qanal276}) is confirmed due to equivalence of the cases $L=0$ and $L = \frac{M}{2}$. A trigonometric transformation of (\ref{qanal278}) demonstrates that $\bigl({\bold\Delta}^K \bigr)_{j m}=0$ at $K-|j-m| = 1 ({\rm mod}\,2)$. $\Box$

{\sf Proposition~4} demonstrates that Ramus's identity allows one to express the entries of ${\bold\Delta}^K$ as the lacunary sums of the binomial coefficients. On another hand, ${\bold\Delta}^K$ obeys (\ref{dmcf}), which is the particular case of
(\ref{mpcf6}) at $h=0$. Therefore it looks appropriate to relate (\ref{mpcf6}) at arbitrary $N$
with generalized Ramus's identities. Regarding (\ref{ratbe7}) and (\ref{qanal272}), we formulate

\vskip0.3cm \noindent {\bf Proposition~5\,} \textsf{(generalized Ramus's identity):\,}
\textit{The following identity is valid}:
\begin{align}
\sum_{|{\bf n}|=K} P({\bf n})\,
{\bold\Delta}^{\bf n}_{{\bmu^L}; {\bmu^R}} &=\,
\displaystyle{\frac{2^{K+N}}{ M^N}} \sum \limits_{{\bf{l}}_N \in {\sf P}^N}
\left(\sum\limits_{k=1}^N \cos \Bigl(\frac{2\pi}{M} {l_k}\Bigr) \right)^K \nonumber \\
&\times\,\prod_{s=1}^{N}
\cos\Bigl(\frac{2\pi}{M} {l_s}
({\mu_{s}^L}-{\mu_s^R}) \Bigr) \,,
\label{ratbeco1}
\end{align}
\textit{where}
\begin{equation}
\label{qanal27}
{\bold\Delta}^{\bf n}_{{\bmu^L}; {\bmu^R}} \equiv \prod_{j=1}^{N}  ({\bold\Delta}^{n_j} )_{\mu^L_{j}; \mu^R_j} \,,
\end{equation}
\textit{$({\bold \Delta}^{n})_{j m}$ is defined by \eqref{qanal276},  \eqref{qanal27611},
and $({\bold \Delta}^{0})_{j m} = {\delta}_{j m}$. Summation indices ${n_j}$ in left-hand side
of \eqref{ratbeco1} are of the same parity as $|\mu^L_{j}- \mu^R_j|$, $1\le j\le N$. Summation in right-hand side of \eqref{ratbeco1} is over $N$-tuples ${\bf l}_N = ({l_1},
{l_2}, \ldots, {l_N})$, $l_k \in {\sf P}\equiv \{0, 1, \ldots, \frac{M}{2}-1\}$}.

\vskip0.3cm \noindent
\noindent{\bf Proof:\,}
Equation (\ref{ratbeco1}) is reduced at $N=1$ to Ramus's identity \eqref{rams}, and it is directly verified at $N=2$.
Mathematical induction with respect to $N$ is straightforward
(Ramus's identity is a base case of induction) and relies upon the fact that left-hand side of (\ref{ratbeco1}) is represented at any $1\le m\le N$:
\begin{equation}
\label{mulr}
\sum_{p=0}^{K} \begin{pmatrix}
K \\
p \end{pmatrix}
({\Delta}^{p})_{{\mu^L_m}; {\mu^R_m}} \sum_{|{ {\bf n}}_{N-1}(m)|=K-p} P({  {\bf n}}_{N-1}(m))\,
{\bold\Delta}^{{{\bf n}}_{N-1}(m)}_{
{{\bmu^L_{N-1}}(m)}; {{\bmu^R_{N-1}}(m)}  }\,,
\end{equation}
where ${\bf n}_{N-1}(m) \equiv (n_1, n_2, \ldots, n_{m-1}, n_{m+1}, \ldots, n_N)$, and, analogously,
\begin{align}
\nonumber
{\bmu^\#_{N-1}}(m) \equiv (\mu^\#_1, \mu^\#_2, \ldots, \mu^\#_{m-1}, \mu^\#_{m+1}, \ldots, \mu^\#_N)\,,\quad \#\,\, {\rm {is}}\,\,L\,\, {\rm {or}}\,\, R\,.\qquad\qquad\qquad\Box
\end{align}

\vskip0.3cm \noindent {\bf Corollary:\,}

\noindent\textit{$\bullet\,\,$
Determinantal generalization of \eqref{ratbeco1} reads}:
\begin{align}
\sum_{|{\bf n}|=K} P({\bf n})\,
\det\bigl(({\bold\Delta}^{n_j} )_{\mu^L_{i}; \mu^R_j}\bigr)_{1
\le i, j \le N} &=\,
\displaystyle{\frac{1}{M^N}}
\sum\limits_{\{{\bphi}_N\}}
\Bigr(2\sum\limits_{m=1}^N \cos {\phi}_{m}\Bigl)^K \nonumber \\
&\times\,|{\CV} (e^{i {\bphi}_N})|^2\,
S_{{\blad^L}}(e^{i {\bphi}_N})
S_{{\blad^R}}(e^{-i {\bphi}_N})\,,
\label{ratbeco}
\end{align}
\textit{where the entries $({\bold\Delta}^{n_j} )_{\mu^L_{i}; \mu^R_j}$, $1
\le i, j \le N$, are given by \eqref{qanal276}} (cf. \cite{nest}).

\noindent\textit{$\bullet\,\,$
The Schur polynomials are equal to unity, $S_{{\blad^L}}(e^{i {\bphi}_N})=S_{{\blad^R}}(e^{-i {\bphi}_N})=1$, provided that ${\bmu^L} = {\bmu^R} = \bdl_N$, where $\bdl_N$ is defined by \eqref{strict}. Then, Eq.~\eqref{ratbeco}
gives the number of non-intersecting trajectories of $N$ random turns walkers initially located at $\bdl_N$ and returning to their initial positions after $K$ steps over long enough chain} ($M\gg 1$):
\begin{align}
\label{ratbe66}
&\sum_{|{\bf n}|=K} P({\bf n})\,
\det\left(\begin{pmatrix}{n_j} \\ \frac{{n_j}+j-i}2
\end{pmatrix}\right)_{1
\le i, j \le N}\,=\,2^K
{\cal J} (K, N)\,, \\
\label{ratbe661}
{\cal J} (K, N) &\equiv\,
\displaystyle{\frac{1}{N!}}\,
\int\limits_{-\pi}^{\pi}
\int\limits_{-\pi}^{\pi}\dots
\int\limits_{-\pi}^{\pi}
\Bigr(\sum\limits_{m=1}^N \cos {\phi}_{m}\Bigl)^K |{\CV} (e^{i {\bphi}_N})|^2\,\frac{d\phi_1 d\phi_2\dots
d\phi_N}{(2\pi)^N}\,,
\end{align}
\textit{where zero values are assigned to the entries of the matrix in \eqref{ratbe66} provided that $n_j$ and $|i-j|$ are of opposite parity. Moreover, when $n_j$ vanishes at some $j$, the entry of the matrix
is Kronecker symbol $\dl_{i j}$}.

\vskip0.3cm The integral ${\cal J} (K, N)$ (\ref{ratbe661}) is zero for $K$ odd (the same is true for the series on the left-hand side of (\ref{ratbe66})), whereas ${\cal J} (K, N)$ is related at $K$ even with the number of
random permutations of $\{1, \ldots, \frac{K}{2}\}$ with at most $N$ increasing subsequences
\cite{forr}, as well as with the distribution of the length of the longest increasing subsequence of random permutations of $\{1, \ldots, \frac{K}{2}\}$ \cite{joh, joh1}. The problem of the longest increasing subsequence of random permutations is related to the random unitary matrices \cite{rain}, whereas more on connection of the longest increasing subsequence with various areas of mathematics can be found in \cite{dan}.

\subsection{Transition amplitude as the generating function of random turns walks}
\label{s333}

Connection between the solutions to (\ref{mpcf3}) in the determinantal form (\ref{ratbe7}) and the series form (\ref{new1}) (with coefficients given by {\sf Proposition~3}) is
expressed by

\vskip0.3cm \noindent {\bf Proposition~6:\,}
\textit{The determinantal
solution $G^0_{{\bmu^L}; {\bmu^R}}(\be)$ is the generating function of the numbers $|P^0_K ({\bmu^L_N} \rightarrow {\bmu^R_N})|$ given by {\sf Proposition~3} followed by \eqref{qanal276}, \eqref{qanal27611}} (\textit{a kind of generalized Ramus's identity}):
\begin{equation}
\label{ratbe66212}
\mathcal D^K_{\be/2}\, \displaystyle{
\det \bigl(
{G}^0_{ {\mu_{n}^L}; {\mu_{k}^R}} (\be) \bigr)_{1\le
n, k \le N}}\,=\,
|P^0_K ({\bmu^L_N} \rightarrow {\bmu^R_N})|\,.
\end{equation}

\vskip0.3cm \noindent
\noindent{\bf Proof:\,}
The formula of differentiation of product of functions enables to obtain \eqref{ratbe66212}
provided that the definition of the determinant is used. In turn, the proof of \eqref{ratbe66212} by induction includes the base case given by \eqref{new1}, \eqref{qanal13}, and \eqref{avv}.
Let us introduce the minor ${\sf G}^{k p}$ labeled by $1\le k, p\le N$ as the determinant of $(N-1) \times (N-1)$ matrix $\{{G}^0_{ {\mu_{i}^L}; {\mu_{j}^R}}\}_{1\le i, j\le N, i\ne k, j\ne p}$ with the entries given by \eqref{qanal277}, and the corresponding cofactor is ${\sf A}_{k p} \equiv (-1)^{k+p} {\sf G}^{k p}$. Left-hand side of \eqref{ratbe66212} acquires the following form when the determinant is expanded along $N^{\rm th}$ column:
\begin{equation}
\label{ratbe66621}
\sum_{j=1}^{N}
\sum_{m_1+m_2=K}\, P(m_1, m_2)\,
({\Delta}^{m_1})_{{\mu^L_{j}}; {\mu^R_{N}}}
\mathcal D^{m_2}_{\be/2}\, {\sf A}_{j N} \,,
\end{equation}
where \eqref{qanal277} and \eqref{avv} are used. Assuming the validity of \eqref{ratbe66212} at
$N\mapsto N-1$ in order to express $\mathcal D^{m_2}_{\be/2}\, {\sf A}_{j N}$, one concludes that \eqref{ratbe66212}
is also valid.
$\Box$

When $M$ is large enough, Eq.~\eqref{ratbe66212} is approximately reduced to
\begin{equation}
\label{ratbe666212}
\mathcal D^K_{\be/2}\det\bigl(I_{|{ {\mu_{i}^L}- {\mu_{j}^R}}|}(\be)\bigr)_{1
\le i, j \le N}\,=\,
|P^0_K ({\bmu^L_N} \rightarrow {\bmu^R_N})|\,,
\end{equation}
where \eqref{qanal276}, \eqref{qanal27611} at $n=0$ are used in $|P^0_K ({\bmu^L_N} \rightarrow {\bmu^R_N})|$ \eqref{qanal272}. Equation (\ref{ratbe666212}) generalizes the case of $N=1$
corresponding to Eqs.~(\ref{25}), (\ref{26}): the Bessel function of the first kind is the generating function of sets of paths between two sites of infinite chain \cite{b1}.

Equation (\ref{ratbe666212}) is specified in the particular case ${\bmu^L_N} = {\bmu^R_N} = \bdl_N$:
\begin{align}
\label{ratbe662}
{\mathcal D}^K_{\be/2}\,
z(2/\be, N)
& =\,|P^0_K ({\bdl}_N \rightarrow\,{\bdl}_N)| \,,\\
\label{ratbe663}
z(2/\be, N) & \equiv\, \det\bigl(I_{|i-j|}(\be)\bigr)_{1
\le i, j \le N}\,,
\end{align}
where $|P^0_K ({\bdl}_N \rightarrow\,{\bdl}_N)|$ is given by left-hand side of  (\ref{ratbe66}), and $z(2/\be, N)$ (\ref{ratbe663}) coincides with the correlation function $G^0_{{\bdl}_N; {\bdl}_N}(\be)$ (\ref{ratbe6}) at large enough $M$. On another hand,
$z(2/\be, N)$ is the Gross-Witten partition function, which demonstrates a third order phase transition at $N\to \infty$, \cite{gross}.

\section{Norm-trace generating function of plane partitions}
\label{sec53}
\subsection{Flipped spins and constrained plane partitions}
\label{sec555}

The commutation relation
\begin{equation}
\label{cor:lin551}
e^{\cQ}\,\si^\pm_{k}\,=\,e^{\mp\al_k} \si^\pm_{k}\,
e^{\cQ}
\end{equation}
and ${\cQ}\mid\Uparrow\rangle = 0$ allow us to obtain
the matrix element of $e^{\cQ}$ over `off-shell' (i.e., arbitrarily parameterized) $N$-particle
states \eqref{conwf1} and \eqref{conj}:
\begin{equation}
\langle \Psi({\bf v}_N)\mid e^{\cQ}\mid \!\Psi({\textbf u}_N)\rangle =
{\cal P}_{\CK}({\textbf v}^{-2}_N, {\textbf
u}^2_N, {\bf a}_M)\,,
\label{cor:liin552}
\end{equation}
where
\begin{equation}
{\cal P}_{\CK}({\textbf v}^{-2}_N, {\textbf
u}^2_N, {\bf a}_M)
\equiv
\sum_{\blad \subseteq \{{\CK}^N\}}S_\blad
({\textbf v}^{-2}_N)S_\blad ({\textbf
u}^2_N) \prod_{i=1}^N e^{\al_{\mu_i}}
\label{cor:lin552}
\end{equation}
is the sum parameterized by $M$-tuple
${\bf a}_M \equiv (\al_{1},
\al_{2}, \ldots ,
\al_{{M}})$, while the parts of $\blad$ and $\bmu$ are related according to (\ref{eqnpart}).
The generic Cauchy--Binet formula
\cite{gant} leads to

\vskip0.3cm \noindent
\noindent{\bf Proposition~7:\,}
\textit{The sum of the Schur polynomials ${\cal P}_{\CK} ({\bf v}^{-2}_N, {\bf
u}^2_N, {\bf a}_M)$ \eqref{cor:lin552} parameterized by $M$-tuple ${\bf a}_M$
admits the determinantal representation}:
\begin{equation}
{\cal P}_{\CK}({\textbf v}^{-2}_N, {\textbf
u}^2_N, {\bf a}_M)\,=\, \displaystyle{
\frac{1}{{\CV} ({\textbf
u}^2_N){\CV} ({\textbf v}^{-2}_N)} \det
\begin{pmatrix} \displaystyle{ \sum\limits_{n=1}^{M} e^{\al_n}\,
\Bigl(\frac{u_i^{2}}{v_j^{2}} \Bigr)^{n-1} } \end{pmatrix}_{1\le i, j \le N}}\,,
\label{cor:lin553}
\end{equation}
\textit{where the Vandermonde determinant \eqref{spxx1} is used}.

Let us introduce the `tilded' notations to be used below by means of the following
\vskip0.2cm
\noindent{\bf Definition~3:\,}
\textit{Let us fix a strict partition ${\bf k}_l \equiv (k_1, k_2, \ldots, k_l)$ of length $l$, $M\ge k_1 > k_2 > \cdots > k_l \ge 1$ $($clearly, $k_j\ge l-j+1$, $1\le j\le l$$)$. Among all strict partitions ${\bmu}_N$ \eqref{strict2}, there exists a subset of such
partitions $\w{\bmu}_N$ that $l$ their parts are given by ${\bf k}_l$, $l\le N$. We introduce strict partitions ${\bf m}_l \subset {\bdl}_N$, where ${\bdl}_N$ is given by \eqref{strict}, such that
parts of ${\bf m}_l$ label positions of the parts of ${{\bf k}}_l$ in $\w{\bmu}_N$. Therefore, non-strict partitions $\w{\blad}\equiv \w{\blad}_N = \w{\bmu}_N - {\bdl}_N$ are characterized by $l$ parts occupying the positions ${{\bf m}}_l$ and given by non-strict partition ${{\bf k}}_l-{\bf m}_l$}.

\vskip0.2cm
\noindent Then, off-shell $N$-particle matrix element of ${\varPi}_{\bf k}$ defined in (\ref{cor:lin611}) arises from (\ref{cor:liin552}):
\begin{align}
\langle \Psi({\bf v}_N)\mid
{\varPi}_{\bf k}
\mid\! \Psi({\textbf u}_N)\rangle &=\,\lim\limits_{{\textbf a}_{_M} \to 0} {\cd}^l_{\al_{k_1} \al_{k_2} \ldots \al_{k_l}}
\nonumber\\
&\times\,\langle \Psi({\bf v}_N)\mid e^{\cQ}\mid\!\Psi({\textbf u}_N)\rangle\,=\,{\w {\cal P}}_{\CK}({\textbf v}^{-2}_N, {\textbf
u}^2_N, {\bf k}_l)\,,
\label{5544}
\end{align}
where the tilded notation ${\w {\cal P}}_{\CK} ({\textbf v}^{-2}_N, {\textbf u}^2_N, {\bf k}_l)$ implies the sum
\begin{equation}
\label{55441}
{\w {\cal P}}_{\CK}({\textbf v}^{-2}_N, {\textbf
u}^2_N, {\bf k}_l)\,\equiv\,
\sum_{\{\w{\blad}\}} S_{\w{\blad}}
({\textbf v}^{-2}_N) S_{\w{\blad}} ({\textbf
u}^2_N)\,,
\end{equation}
and summation in \eqref{55441} goes over ${\w{\blad}}$ with respect to fixed $l$-tuple ${\bf k}_l$ (cf. {\sf Definition~3}).

The matrix element of ${\varPi}_{\bf k}$
under the $q$-parametrization  (\ref{rep21}) arises from (\ref{5544}) and (\ref{55441}):
\begin{equation}
<{\varPi}_{\bf k}>_{N, q}\, \equiv\,\langle \Psi(\textbf{q}^{-1/2}_N)\mid {\varPi}_{\bf k} \mid\!
\Psi((\textbf{q}_N/{q})^{1/2})
\rangle\,=\,{\w {\cal P}}_{\CK}
\Bigl(\textbf{q}_N, \frac{\textbf{q}_N}{q}, {\bf k}_l\Bigr)
\label{cor:lenn5571} \,.
\end{equation}
Equation (\ref{cor:lenn5571}) in the case ${\bf k}_l = {\bdl}_l$ (see \eqref{strict}) reads:
\begin{equation}
<{\prod}_{i=1}^{l} {\sf q}_{i} >_{N, q}\,=\,{\w {\cal P}}_{\CK}
\Bigl(\textbf{q}_N, \frac{\textbf{q}_N}{q}, {\bdl}_l \Bigr)
\label{cor:lin5571} \,,
\end{equation}
so that $\w{\bmu}_N$ and $\w{\blad}_N = \w{\bmu}_N - \bdl_N$ in (\ref{55441}) are concretized:
\begin{align}
\w{\bmu}_N & = (\mu_1, \mu_2, \ldots, \mu_{N-l}, l, l-1, \ldots, 1)\,,
\label{cor:lin558} \\
\w{\blad}_N & = (\la_1, \la_2, \ldots, \la_{N-l}, 0, 0, \ldots, 0)\,,
\label{cor:lin5581}
\end{align}
and summation in ${\w {\cal P}}_{\CK}$ (\ref{55441}) is over $\CK\geq \la_1\geq \la_2\geq \dots\geq
\la_{N-l}\geq 0$.

According to (\ref{schrepr5}), right-hand side of (\ref{cor:lin5571}) provides the generating function of watermelons depicted in Figure~\ref{fig:f6}:
\begin{equation}
\lim_{q\to 1}
<{\prod}_{i=1}^{l} {\sf q}_{i} >_{N, q}\,=\,{\w {\cal P}}_{\CK}
(\textbf{1}_N, \textbf{1}_N, {\bdl}_l) \,=\, \sum_{\{{\w\blad}\}} S_{\w\blad} (\textbf{1}_N)
S_{\w\blad} (\textbf{1}_N)
\label{cor:lin5572} \,.
\end{equation}
Indeed, $S_{\w\blad} ({\bf 1}_N)$ corresponds to the paths  connecting the points $C_i=(i, N+1-i)$ and ($N, \w\mu_i$), where $\w\mu_i$ are given by (\ref{cor:lin558}). The nest in Figure~\ref{fig:f5} is just depicted for $\w\bmu$ (\ref{cor:lin558}) at $l=3$.
An $i^{\rm th}$ path in Figure~\ref{fig:f5} makes $\la_i \in \blad_{N-l}\equiv (\lambda_1, \lambda_2, \ldots, \lambda_{N-l})$ steps upwards
at $1\le i\le N-l$, while only rightward steps are allowed at $N-l+1\le i\le N$.

In the case of $l=0$, Eq.~(\ref{cor:lin5572}) is reduced to (\ref{schrepr5}), and the sum ${\cal P}_{\CK}({\bf 1}_N, {\bf 1}_N, {\bf 0}_{M})$ (\ref{cor:lin552}) provides the number of such watermelons that upward steps are allowed for
all paths from $1^{\rm st}$ to $N^{\rm th}$. The number ${\cal P}_{\CK}({\bf 1}_N, {\bf 1}_N, {\bf 0}_{M})$ is also interpreted
(cf. Figure \ref{fig:ff9}) as the number $A (N, N, M-N)$ of plane partitions in $N\times N \times (M-N)$ box (MacMahon formula, \cite{bres}):
\begin{equation}
\label{ratbe794}
{\cal P}_{\CK}({\bf 1}_N, {\bf 1}_N, {\bf 0}_{M})
= A (N, N, M-N) = \prod_{k=1}^{N} \prod_{j=1}^{N} \frac{M-N+k+j-1}{k+j-1}\,.
\end{equation}

As far as the bijection between the watermelons and the plane partitions is concerned, the watermelons characterized by $\w{\bmu}_N$ (\ref{cor:lin558}) and $\w{\blad}_N$ (\ref{cor:lin5581}) are mapped to such stacks of cubes that $l\times l$ square remains empty on the bottom of $N\times N\times {\CM}$ box. Indeed, the watermelon in Figure \ref{fig:f6} is characterized by
$\mu_4=3$, $\mu_5=2$, $\mu_6=1$, and $\la_4=\la_5=\la_6=0$. Three columns of zero height on the diagonal of the plane partition lead to the empty $3\times 3$ square dashed in Figure \ref{fig:ff9}.
Therefore, ${\w {\cal P}}_{\CK}
(\textbf{1}_N, \textbf{1}_N, {\bdl}_l)$ (\ref{cor:lin5572}) enumerates the plane partitions constrained by presence of the empty square.

Generally, ${\w {\cal P}}_{\CK}
(\textbf{1}_N, {\textbf{1}_N}, {\bf k}_l)$ corresponding to (\ref{55441}) enumerates the plane partitions with $l$ columns of heights given by
${{\bf k}}_l-{\bf m}_l$
at the positions labelled by parts of ${\bf m}_l$ (cf. {\sf Definition~3}).

\subsection{Norm-trace generating function}
\label{sec55}

The \textit{norm-trace generating function} ${G}(N, N, {\CM} |\,q, \ga)$, i.e., the generating function of plane partitions with fixed total volume of the parts on principal diagonal in box of height $\CM$ and bottom of size $N\times N$, has been derived in \cite{st} and generalized in \cite{gan}. The determinantal representation for  ${G}(N, N, {\CM} |\,q, \ga)$ has been derived for the model of strongly correlated bosons
\cite{statm}. The determinantal formula for the generating function of plane partitions with fixed heights of several diagonals has been obtained by means of the four-vertex model in inhomogeneous field \cite{bmjpa}.

The norm-trace generating function ${G}(N, N, {\CM} |\,q, \ga)$ arises from Eqs.~(\ref{cor:liin552}) and (\ref{cor:lin552})
under the $q$-parametrization (\ref{rep21}). Indeed, let us
consider the linear parametrization of ${\bf a}_M$ and specify $\al_n$ so that $e^{\al_n} = \ga^n$, $0<{\ga}\le 1$. Using $<e^{\cQ (\ga)} >_{N, q}$ to denote the $q$-parameterized matrix element (\ref{cor:liin552}),
\begin{equation}
<e^{\cQ (\ga)}>_{N,q}\,\equiv\,\langle \Psi(\textbf{q}^{-1/2}_N)\mid e^{\cQ (\ga)}\mid\!
\Psi((\textbf{q}_N/{q})^{1/2})
\rangle\,,
\label{ratbee707}
\end{equation}
one formulates the following

\vskip0.3cm \noindent
\noindent{\bf Proposition~8:\,}
\textit{The determinantal representation for
the norm-trace generating function of plane partitions with
fixed total volume of the parts on principal diagonal in a box of height $\CM$ and $N\times N$ bottom is given}:
\begin{align}
{G}(N, N, {\CM} |\,q, \ga) & =
\ga^{\frac{-N}{2}(N+1)}\,<e^{\cQ (\ga)}>_{N,q}
\nonumber \\
&=\,\displaystyle{
\frac{\det\big(
{\sf h}_M( {\ga}\, q^{i+j-1})\big)_{1\le i, j \le N}}{{\CV} (\textbf{q}_N/{q}) {\CV} (\ga\,\textbf{q}_N)}}\,,
\label{ratbe707}
\end{align}
\textit{where ${\sf h}_M$ is defined by} (\ref{cauchy1}).

\vskip0.3cm \noindent
\noindent{\bf Proof:\,}
First of all, one obtains from (\ref{cor:lin552}) and (\ref{cor:lin553}):
\begin{align}
\ga^{\frac{-N}{2}(N+1)}\,
<e^{\cQ (\ga)}>_{N,q} &=\,
\sum_{\blad \subseteq \{{\CK}^N\}} \gamma^{|{\blad}|} S_{\blad} \Bigl(\frac{\textbf{q}}{q}\Bigr)
S_{\blad} (\textbf{q})
\label{cor:lin559} \\
&=\,\displaystyle{
\frac{1}{{\CV} (\textbf{q}_N/{q}) {\CV} (\ga\,\textbf{q}_N)} \det
\begin{pmatrix} \displaystyle{ \sum\limits_{n=0}^{M-1} \ga^{n}
q^{n(i+j-1)} } \end{pmatrix}_{1\le i, j \le N}}\,,
\label{cor:lin5591}
\end{align}
where $|\blad|$ is the volume of $\blad$, and the homogeneity property $\ga^{|\blad|} S_{\blad} (\textbf{q}) = S_{\blad} (\ga\,\textbf{q})$ is used to pass from (\ref{cor:lin559}) to (\ref{cor:lin5591}).
The series on right-hand side of (\ref{cor:lin559}) is the norm-trace generating function of plane partitions with fixed height of their diagonal parts in $N\times N\times \CM$ box. Therefore,
the statement (\ref{ratbe707}) for ${G}(N, N, {\CM} |\,q, \ga)$ is valid because of the determinanal formula (\ref{cor:lin5591}), \cite{statm}. Equation (\ref{ratbe707}) at ${\ga}=1$ gives the determinantal formula for the generating function of boxed plane partitions in $N\times N\times {\CM}$ box:
\begin{equation}
\label{cor:llin5591}
\lim_{q\to 1}\,{G}(N, N, {\CM} |\,q, 1) = A (N, N, {\CM}) \,,
\end{equation}
where $A (N, N, {\CM})$ is the number of plane partitions (\ref{ratbe794}). $\Box$

Assume that the approximation ${\sf h}_M(x) \simeq (1-x)^{-1}$ is valid at $|x|< 1$ and large enough $M$. Then, one obtains from (\ref{ratbe707}):
\begin{align}
\lim_{M\to\infty}
{G}(N, N, {\CM} |\,q, \ga) & =\,
\frac{\det \begin{pmatrix} \bigl(1- \ga
q^{i+j-1}\bigr)^{-1} \end{pmatrix}_{1\le i, j \le N}}
{{\CV} (\textbf{q}_N/{q}) {\CV} (\ga\,\textbf{q}_N)}
\label{cor:lin5596}
\\
&=\,\displaystyle{
\prod_{i=1}^{N} \prod_{j=1}^{N} \frac{1}{1- \ga
q^{i+j-1} }}\,.
\label{cor:lin5597}
\end{align}
Evaluation of the
Cauchy-type determinant on right-hand side of (\ref{cor:lin5596}) leads to the double product  (\ref{cor:lin5597}), which is nothing but the norm-trace generating function of plane partitions with unbounded height \cite{statm}. Further, one obtains from (\ref{cor:lin5597}) in the double scaling limit:
\begin{equation}
\label{cor:lin5598}
\lim_{N/M\ll 1, N \to\infty}
{G}(N, N, {\CM} |\,q, \ga) \,=\,\displaystyle{
\prod_{n=1}^{\infty}  \frac{1}{(1- \ga
q^{n})^n }}\,.
\end{equation}
Right-hand side of (\ref{cor:lin5598}) coincides with the expression provided in \cite{jap} as the partition function of the five-dimensional supersymmetric Yang-Mills theory.
Two-dimensional
Conformal Field Theory is referred in \cite{jap} as a way of derivation of the partition function.

\section{Matrix elements of
the generating exponential on the Bethe states and the lattice paths}
\label{ss342}

Let us begin with the calculation of the normalized matrix element of the generating exponential over the on-shell Bethe states ({\sf Proposition~1}):
\begin{equation}
\langle e^{\cal Q}\rangle_N \equiv \frac{\langle \Psi (e^{i{\bth}_N/2}) \!\mid
e^{\cal Q} \mid\! \Psi (e^{i{\bth}_N/2}) \rangle}{\CN^2 (e^{i{\bth}_N/2})}\,, \label{ratbe0}
\end{equation}
where ${\cal Q}$ is given by (\ref{cor:lin05}), and $N$-tuple $e^{{i\bth}_N/2} = (e^{i\ta_{1}/2}, e^{i\ta_{2}/2}, \ldots , e^{i\ta_{N}/2})$ is to express
the substitute $v_j = u_j\equiv
e^{i\ta_{j}/2}$ ($1\le j\le N$).
Using {\sf Proposition~7} and the Bethe solution (\ref{besol}), we express $\langle e^{\cal Q}\rangle_N$ (\ref{ratbe0}):
\begin{equation}
\langle e^{\cal Q}\rangle_N =
\det(e^{\widehat\al})\,,\qquad
e^{\widehat\al} \equiv \Bigl(
\displaystyle{ \frac{1}{M}\sum\limits_{n=1}^{M} e^{\al_n + i n (\theta_i-\theta_j)}} \Bigr)_{1\le i,j \le N}\,.
\label{cor:lin55311}
\end{equation}
Under the conventional
choice of ${\bf a}_M$, the matrix element $\langle e^{\al Q(m)}\rangle_N$ (\ref{cor:lin55311}) is known as the generating function of equal-time mean values of third components of spins \cite{vk1, ess, ml3, col, col1}.

Let us obtain the Boltzmann-weighted matrix element of $e^{\cQ}$ with respect to on-shell Bethe state-vectors.
A determinantal expression for the corresponding off-shell average is calculated by insertion of the
decomposition of unity (\ref{field7}):
\begin{align}
\nonumber
& \langle \Psi({\bf v}_N) |
e^{\cQ}\,e^{-\be H} | \Psi({\textbf u}_N)\rangle\,=\,\sum \limits_{\{{\bphi}_N\}}
\frac{e^{-\be E_N ({\bphi}_N)}}{\CN^{2} (e^{i{\bphi}_N/2})} \\
& \times\,\langle \Psi({\bf v}_N) | e^{\cQ} | \Psi(e^{i{\bphi}_N/2})\rangle \,\frac{\det T_M(e^{-i{\bphi}_N}, {\textbf u}^2_N)}{
{\CV} (e^{-i{\bphi}_N}) {\CV} ({\textbf u}^2_N)}\,,
\label{cor:lin661}
\end{align}
where $E_N ({\bphi}_N)$ is given by (\ref{egen}), and (\ref{spxx3}), (\ref{cauchy}),  (\ref{cauchy1}), (\ref{spxx}), (\ref{ratbe91}) are accounted for. Taking into account {\sf Proposition~7} to express $\langle \Psi({\bf v}_N) | e^{\cQ} | \Psi(e^{i{\bphi}_N/2})\rangle$, one obtains:
\begin{align}
\nonumber
& \langle \Psi({\bf v}_N) |
e^{\cQ}\, e^{-\be H} | \Psi({\textbf u}_N)\rangle\,=\, \\
& =\,\displaystyle{ \frac{e^{\be h M/2}}{{\CV}({\textbf
u}_N^2){\CV}({\textbf v}_N^{-2})} \det
\left(\sum\limits_{k, l=1}^{M}
e^{{\al}_{_k}}
{G}_{k;\,l}(\be)\, \frac{u_i^{2 l}}{v_j^{2k}} \right)_{1\le i,j \le N}}\,,
\label{cor:lin663}
\end{align}
where
\begin{equation}
\label{qanal27777}
{G}_{k;\,l} (\be) \,\equiv\, \displaystyle{
\frac{1}{M} \sum\limits_{p\in {\sf S}^\pm}
e^{-\be\ep (p)}\,
e^{i p (l-k)}}\,,\qquad {\sf S}^{\pm} \equiv \bigl\{p \,\big | \cos Mp = (-1)^{N-1} \bigr\}\,,
\end{equation}
and the choice of ${\sf S}^+$ or ${\sf S}^-$ is due $N$ even or odd, respectively.

Equation (\ref{cor:lin663}) on the on-shell Bethe states leads to the normalized matrix element:
\begin{align}
\label{cor:lin664}
\langle e^{\cQ}\, e^{-\be H} \rangle_N & \equiv \frac{\langle \Psi (e^{i{\bth}_N/2}) | e^{\cQ}\, e^{-\be H} | \Psi (e^{i{\bth}_N/2}) \rangle}{\CN^2 (e^{i{\bth}_N/2})} \\
 & =\,e^{\be h M /2}\,\det
\bigl( e^{- \be {\h\ep}} {e^{\h\al}}\bigr)\,,
\label{cor:lin665}
\end{align}
where $e^{\h\al}$ is defined by (\ref{cor:lin55311}), and the diagonal matrix
${\h\ep}$ consists of
$\ep(\ta_j)$ (\ref{egen}):
\begin{equation}
\label{mpcf1}
{\h\ep}\equiv \underset{1\le j\le N}{\diag} \{\ep(\ta_j)\}\,.
\end{equation}

The commutation relation (\ref{cor:lin551}) together with Eqs.~(\ref{mpcf}), (\ref{ratbe7}) allows us to obtain $\langle e^{\cal Q}\, e^{-\be H}\rangle_N$
in the integral form at $M\gg 1$:
\begin{align}
\nonumber
\langle e^{\cal Q}\, e^{-\be H}\rangle_N
\,\simeq\,
\frac{e^{\be h M/2}}{{\CN}^2 (e^{i{\bth}_N/2}) N!}\,\int\limits_{{\sf S}^{N}}
{\cal P}_{\CK}(e^{-i{\bf p}_N}, e^{i{\bth}_N}, {\bf 0}_M)
\\
\times \,{\cal P}_{\CK}
(e^{-i{\bth}_N}, e^{i{\bf p}_N}, {\bf a}_M)\,| \CV (e^{i{\bf p}_N}) |^2\,
e^{{\be}\sum_{l=1}^N
(\cos{p}_l-h)}
\frac{d^Np}{(2\pi)^N}\,,
\label{cor:lin667}
\end{align}
where ${\bf p}_N\equiv (p_1,
p_2, \dots ,
p_N)$, $d^Np = d p_1 d p_2\cdots
d p_N$, and the integration domain ${\sf S}^{N}$ is $N$-fold product
of ${\sf S}\equiv [-\pi, \pi ]$. Provided that the formula (\ref{ratbe7}) for the transition amplitude $G_{{\bmu^L}; {\bmu^R}} (\be)$ is used (\textsf{Proposition~2}), the representation (\ref{cor:lin667}) takes equivalent form:
\begin{align}
\langle e^{\cal Q}\, e^{-\be H}\rangle_N
\,\simeq\,{\CN}^{-2} (e^{i{\bth}_N/2})  \sum\limits_{{\blad^{L, R}}
\subseteq \{\CK^N\}}
S_{{\blad^L}}(e^{-i{\bth}_N})\,
S_{{\blad^R}}(e^{i{\bth}_N})
& \nonumber
\\ \times\,
\exp\Bigl(\sum_{k=1}^{N}\al_{\mu^L_k}
\Bigr)\,
G_{{\bmu^L}; {\bmu^R}} (\be)& \,,
\label{cor:lin668}
\end{align}
where $G_{{\bmu^L}; {\bmu^R}} (\be)$ is expressed at
$M\gg 1$ (see (\ref{ratbe6}), (\ref{qanal277}), and (\ref{ratbe7373})):
\begin{equation}
G_{{\bmu^L}; {\bmu^R}} (\be) \,\simeq \,e^{\be h (\frac{M}2 - N)} \det\bigl(I_{|{ {\mu_{i}^L}- {\mu_{j}^R}}|}(\be)\bigr)_{1
\le i, j \le N}\,.
\label{cor:lin668888}
\end{equation}

As it follows from \textsf{Proposition~6}, the matrix element in the numerator of (\ref{cor:lin664}) just represented by right-hand side of (\ref{cor:lin668}) (clearly, without ${\CN}^{-2} (e^{i{\bth}_N/2})$) is related to
enumeration of lattice walks \cite{nest}. Indeed,
applying $\lim\limits_{{\textbf a}_{_M} \to 0} {\cd}^l_{\al_{1} \al_{2} \ldots \al_{l}}$ (\ref{cor:lin61})
to the numerator of (\ref{cor:lin664})
taken over the ground state solution (\ref{grstxx}), one obtains with the use of
(\ref{cor:lin668}) and \eqref{cor:lin668888}:
\begin{equation}
\mathcal D^K_{\be/2}\,
\langle \Psi (e^{i{\bth}^{\rm g}_N/2}) \Bigl| {\prod}_{i=1}^{l} {\sf q}_{i}\, e^{- \be H} \Bigr|
\Psi (e^{i{\bth}^{\rm g}_N/2})
\rangle\,=\, \mathfrak{P}(e^{i{\bth}^{\rm g}_N/2}; e^{i{\bth}^{\rm g}_N/2}\,| K)\,,
\label{cor:linn5571}
\end{equation}
where $\mathfrak{P}(e^{i{\bth}^{\rm g}_N/2}; e^{i{\bth}^{\rm g}_N/2}\,| K)$ is the value of the polynomial
\begin{equation}
\label{cor:linnn5571}
\mathfrak{P}({\textbf v}_N; {\textbf u}_N\,| K) \equiv
\sum_{\w\blad^L, \blad^R } S_{\w\blad^L} ({\textbf v}_N^{-2}) S_{\blad^R} ({\textbf u}_N^{2})\,
\mathfrak{G}({\w\bmu^L}; {\bmu^R}\,| K)
\end{equation}
at ${\textbf u}_N = {\textbf v}_N = e^{i{\bth}^{\rm g}_N/2}$.
Summation in \eqref{cor:linnn5571} over ${\blad}^R$
and $\w\blad^L$ is according to (\ref{strict1}) and (\ref{cor:lin5581}), respectively, while
$\mathfrak{G}({\w\bmu^L}; {\bmu^R}\,| K)$ is given by (\ref{mpcf4}).

The replacement
$e^{i{\bth}^{\rm g}_N}\longmapsto 1$ is appropriate at $M\gg N$, and one obtains from (\ref{cor:linn5571}):
\begin{equation}
\lim_{q\to 1} \mathcal D^K_{\be/2} <{\prod}_{i=1}^{l} {\sf q}_{i}\, e^{- \be H} >_{N, q}\,=\, \mathfrak{P}(\textbf{1}_N; \textbf{1}_N | K)\,,
\label{cor:linnn05571}
\end{equation}
where $<\cdot >_{N, q}$ is defined in (\ref{ratbee707}).
Right-hand side of (\ref{cor:linnn05571}) is expressed:
\begin{equation}
\label{cor:linnn15571}
\mathfrak{P}(\textbf{1}_N; \textbf{1}_N
| K) = \sum_{p=0}^{K}
\begin{pmatrix}
K\\
p
\end{pmatrix}
\bigl(h(M-2N) \bigr)^p\,\mathfrak{P}^0 (\textbf{1}_N; \textbf{1}_N
| K-p) \,,
\end{equation}
where $\mathfrak{P}^0 (\textbf{1}_N; \textbf{1}_N
| K-p)$ corresponds to \eqref{cor:linnn5571} expressed by means of $\mathfrak{G}^0 ({\w\bmu^L}; {\bmu^R}\,| \cdot)$,
\begin{equation}
\label{cor:linnn15572}
\mathfrak{P}^0 (\textbf{1}_N; \textbf{1}_N
| K-p) =
\sum_{\w\blad^L, \blad^R } S_{\w\blad^L} (\textbf{1}_N)
S_{\blad^R} (\textbf{1}_N)\,
|P^0_{K-p} ({\w\bmu^L} \rightarrow\,{\bmu^R})|\,,
\end{equation}
and $|P^0_{K-p} ({\w\bmu^L} \rightarrow\,{\bmu^R})|$ is given by {\sf Proposition~3}.

\begin{figure}[h]
\center
\includegraphics [scale=0.8] {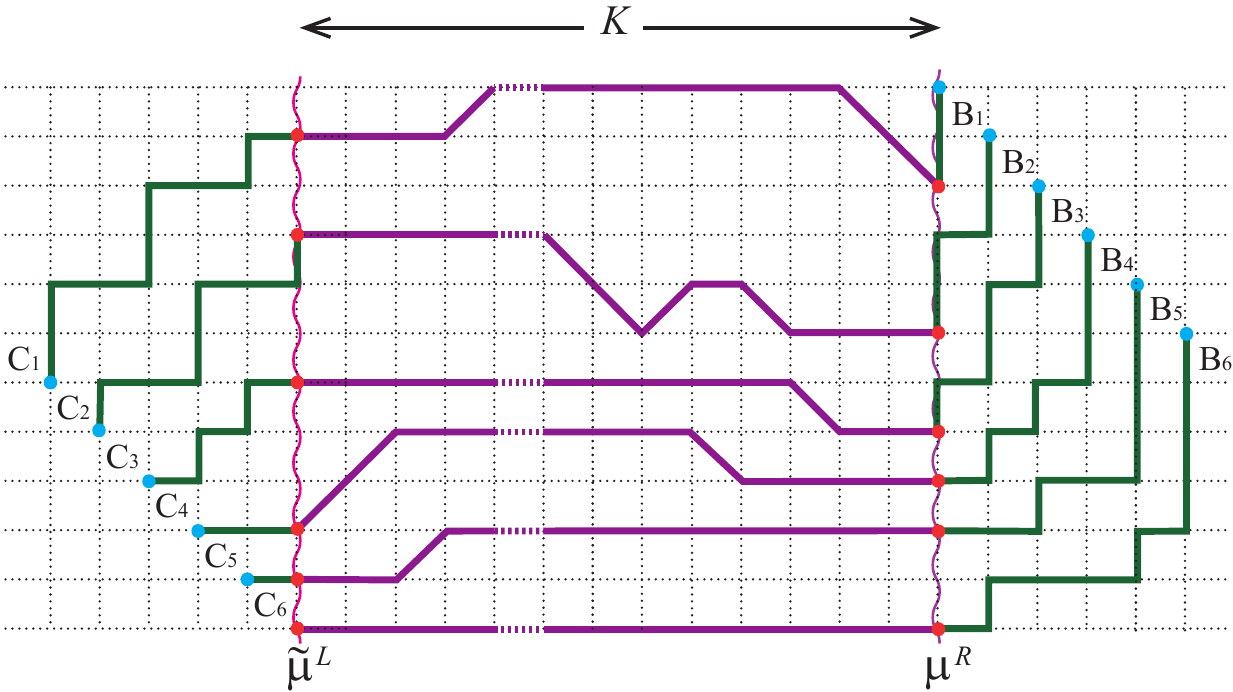}
\caption{Nest of paths contributing to $\mathfrak{P}(\textbf{1}_N; \textbf{1}_N | K)$ at $N=6$, $K=13$, and $p=1$. }
\label{fig:f9}
\end{figure}

The coefficient $\mathfrak{P}^0(\textbf{1}_N; \textbf{1}_N
| K-p)$ (\ref{cor:linnn15571})
enumerates, according to (\ref{cor:linnn15572}), the compound paths. Indeed, the factor $S_{\w\blad^L} (\textbf{1}_N)$ in
(\ref{cor:linnn15572}) corresponds to walks by the lock steps rules from $C_i$, $1\le i\le N$, to the sites ${\w\bmu^L}$ (Figure \ref{fig:f5}). The
contribution $|P^0_{K-p} ({\w\bmu^L} \rightarrow\, {\bmu^R})|$ in
(\ref{cor:linnn15572}) corresponds to the random turns walks from ${\w\bmu^L}$ to ${\bmu^R}$. The factor $S_{\blad^R} (\textbf{1}_N)$ accounts for the lock steps
walks from ${\bmu}^R$ to $B_i$, $1\le i\le N$. The coefficients at powers of $h(M-2N)$ in
$\mathfrak{P}(\textbf{1}_N; \textbf{1}_N
| K)$ (\ref{cor:linnn15571}) are
responsible for enumeration of non-intersecting lattice walks corresponding to \eqref{cor:linnn15572} but with stays inserted. Therefore
(\ref{cor:linn5571}) and (\ref{cor:linnn05571})
demonstrate that the matrix element of ${e^{\cQ}\,e^{-\be H} }$ over on-shell Bethe states
is the generating function of numbers of nests of non-intersecting paths
of the type presented in Figure~\ref{fig:f9}. A typical nest is depicted in Figure~\ref{fig:f9} for $K=13$ and $p=1$, so that the paths are characterized by ${\bf n}=(0, 1, 3, 1, 4, 3)$, $|{\bf n}|= 12$.

\section{Constrained lattice paths and plane partitions}
\label{sec60}

\subsection{The $N$-particle mean values and constrained lattice paths}

Let ${\sf trace}$ in (\ref{cor:lin5}) to imply summation over all $N$-particle Bethe solutions to (\ref{betheexp}), and let us consider the $N$-particle trace of the Boltzmann-weighted generating exponential:
\begin{equation}
\label{ratbee7003}
\tr_N(e^{\cQ} e^{-\be H})\,\equiv\,
\sum_{\{{\bth}_N \}} \langle e^{\cQ}\, e^{-\be H} \rangle_N\,.
\end{equation}
The definition (\ref{ratbee7003})
leads to $N$-particle mean value dependent on the parameter $\be$: \begin{equation}
G_{N, \be}\, \equiv\,
{{\langle\langle}} e^{\cQ} {{\rangle\rangle}}_{N, \be}\,
\equiv\, \tr_N(e^{\cQ}\boldsymbol{\rho}_N )\,,\qquad \boldsymbol{\rho}_N \equiv \frac{e^{-\be H}}{\tr_N(e^{-\be H})}
\,.
\label{new11}
\end{equation}

In order to investigate (\ref{new11}) at $M\gg 1$,
let us express (\ref{ratbee7003}) using the representation (\ref{cor:lin668}), (\ref{cor:lin668888}):
\begin{equation}
\tr_N(e^{\cQ} e^{-\be H}) = \sum\limits_{\{{\bmu}_N\}}
\exp\Bigl(\sum_{k=1}^{N}\al_{\mu_k}
\Bigr)\,G_{{\bmu}; {\bmu}} (\be) \label{ratbe7003}\,,
\end{equation}
where (\ref{ratbe77}) enables to sum up over the sets of the Bethe solutions. The transition amplitude $G_{{\bmu}^L; {\bmu}^R} (\be)$ (defined by \eqref{mpcf}) is related with enumeration of random walks of $N$ vicious walkers with initial/final
positions given by partitions ${\bmu}^L$ and ${\bmu}^R$ (Section~\ref{50}).
Only closed walks contribute to
$\tr_N(e^{\cQ} e^{-\be H})$ \eqref{ratbe7003}, since the summands in right-hand side contain the diagonal entries $G_{{\bmu}; {\bmu}} (\be)$.

\noindent $\bullet\,\,$ The mean value (\ref{new11}) is estimated with the use of \eqref{cor:lin668888}
in (\ref{ratbe7003}):
\begin{equation}
G_{N, \be} ({\bf a}_{M})\Big|_{M\gg 1}\,
\simeq\,\frac{{\cal I}_N (\be, {\bf a}_{M})}{{\cal I}_N (\be, {\bf 0}_{M})} \,,
\label{new12}
\end{equation}
where
\begin{equation}
\label{ratbe711432}
{\cal I}_N (\be, {\bf a}_{M}) \equiv
\sum_{\{{\bmu}_N\}} \det \bigl(e^{\al_{\mu_i}} I_{|\mu_i-\mu_j|}(\be) \bigr)_{1\le i, j\le N}\,.
\end{equation}

Let us apply $\mathcal D^l_{\al}$ to (\ref{new12}) under the parametrization ${\bf a}_M^\al \equiv
\frac{\al}N (1, 2, \ldots, M)$  (see (\ref{new30})). In this case, ${\cal I}_N (\be, {\bf a}_{M})$ (\ref{ratbe711432}) takes the form:
\begin{equation}
\label{new15}
{\cal I}_N \bigl(\be, {\bf a}_M^\al \bigr)\,=\,
\sum_{\{{\bmu}_N\}}
e^{\frac{\al}N |{\bmu}|} \det \bigl( I_{|\mu_i-\mu_j|}(\be) \bigr)_{1\le i, j\le N}\,,
\end{equation}
and we obtain using (\ref{new15}) in \eqref{new12}:
\begin{equation}
{{\langle\langle}} {\sf M}^l {{\rangle\rangle}}_{N, \be}\Big|_{M\gg 1}\,
\simeq\,\frac{N^{-l} \sum_{\{{\bmu}_N\}}
|{\bmu}_N|^l \det \bigl( I_{|\mu_i-\mu_j|}(\be) \bigr)_{1\le i, j\le N}}{{\cal I}_N (\be, {\bf 0}_{M})} \,,
\label{new17}
\end{equation}
where ${\sf M}$ is defined in (\ref{new30}).

The following identity is valid for ${\cal I}_N (\be, {\bf a}_{M})$
\eqref{ratbe711432}
due to the generalized Ramus's identity (\ref{ratbe666212})
({\sf Proposition~6}):
\begin{equation}
\label{ratbe6621222}
\mathcal D^K_{\be/2}\, {\cal I}_N (\be, {\bf a}_{M})\,=\,{\cal P}_{K, N} ({\bf a}_{M})\,,
\end{equation}
where
\begin{equation}
\label{ratbe71131}
{\cal P}_{K, N} ({\bf a}_{M})\, \equiv\, \sum_{\{{\bmu}_N\}} \sum_{|{\bf n}|=K} P({\bf n})\,\det\left(
e^{\al_{\mu_j} } \begin{pmatrix}{n_j} \\ \frac{{n_j}+\mu_j-\mu_i}2
\end{pmatrix}\right)_{1
\le i, j \le N}\,.
\end{equation}
Therefore, Eq.~(\ref{ratbe6621222})
under the parametrization ${\bf a}_M^\al$ enables the generating function:
\begin{align}
\nonumber
&\mathcal D^l_{\al}\,\mathcal D^K_{\be/2}\,\Bigl[
e^{\be h (\frac{M}2 - N)} {\cal I}_N \bigl(\be, {\bf a}_M^\al\bigr) \Bigr] \\
&=\, \sum_{p=0}^{K}
\begin{pmatrix}
K\\
p
\end{pmatrix}
\bigl(h(M-2N) \bigr)^p
\mathcal D^l_{\al}\, {\cal P}_{K-p, N} \bigl({\bf a}_M^\al \bigr)
\,,
\label{new16}
\end{align}
where the application of $\mathcal D^l_{\al}$ to ${\cal P}_{K-p, N} ( \cdot)$  with the use of (\ref{new15}) gives
\begin{equation}
{\mathcal D}^l_{\al}\,{\cal P}_{K-p, N} \bigl({\bf a}_M^\al \bigr) =
\frac{1}{N^l} \sum_{\{{\bmu}_N\}} |{\bmu}_N|^l |P^0_{K-p} ({\bmu}_N \rightarrow\,{\bmu}_N)|\,,
\label{new116}
\end{equation}
and the numbers of nests of paths $|P^0_{K-p} ({\bmu}_N \rightarrow\,{\bmu}_N)|$ are:
\begin{equation}
\label{new106}
|P^0_{K-p} ({\bmu}_N \rightarrow\,{\bmu}_N)|=\sum_{|{\bf n}|=K-p} P({\bf n})\,\det\left(
\begin{pmatrix}{n_j} \\ \frac{{n_j}+\mu_j-\mu_i}2
\end{pmatrix}\right)_{1
\le i, j \le N}\,.
\end{equation}

Summation over ${\bmu}_N$
is split into summation over partitions of fixed weight, $|{\bmu}_N|= m$, and over corresponding finite set of $m\in\BN$. It is seen from (\ref{new116}) that right-hand side of (\ref{new16}) is a polynomial with respect to the products $(\frac{m}N)^l (h(M-2N))^p$ possessing the integer coefficients:
\begin{equation}
\label{new18}
\begin{pmatrix}
K\\
p
\end{pmatrix}
\sum_{{\bmu}_N \vdash m} |P^0_{K-p} ({\bmu}_N \rightarrow\,{\bmu}_N)|\,.
\end{equation}
Equation (\ref{new18}) enumerates
the nests of $K$-step closed trajectories with
$p$ stays (compare with \eqref{cor:linnn15571}, \eqref{cor:linnn15572}, and Figure~\ref{fig:f1}). The initial/final pos\-it\-ions of vicious walkers constitute partitions of appropriate $m\in\BN$.

\noindent $\bullet\,\,$ With regard at (\ref{cor:lin611}) and (\ref{cor:lin61}), we obtain from (\ref{new12}):
\begin{equation}
{{\langle\langle}} {\varPi}_{\bf k}\, {{\rangle\rangle}}_{N, \be} \Big|_{M\gg 1}\,
\simeq\,\frac{\w{\cal I}_N (\be | {\bf k}_l)}{{\cal I}_N (\be, {\bf 0}_{M})} \,,
\label{new13}
\end{equation}
where
\begin{equation}
\w{\cal I}_N (\be | {\bf k}_l)\,
\equiv\,
\sum_{\{{\w\bmu}_N\}} \det \bigl(I_{| \w\mu_i - \w\mu_j|}(\be) \bigr)_{1\le i, j\le N}\,,
\label{ratbe711433}
\end{equation}
and summation is over strict partitions given by {\sf Definition~3} for a fixed $l$-tuple ${\bf k}_l$.

The identity (\ref{ratbe666212}) tells us that $\w{\cal I}_N (\be | {\bf k}_l)$ (\ref{ratbe711433}) is the generating function of numbers $\w{\cal P}_{K, N} ({\bf k}_l)$ of nests of trajectories ({\sf Proposition~6}):
\begin{equation}
\mathcal D^K_{\be/2}\, \w{\cal I}_N (\be | {\bf k}_l)\,
=\,\w{\cal P}_{K, N} ({\bf k}_l)\,,
\label{new14}
\end{equation}
where
\begin{align}
\label{ratbe71143}
\w{\cal P}_{K, N} ({\bf k}_l) &\equiv \sum_{\{{\w\bmu}_N\}} |P^0_{K} ({\w\bmu}_N \rightarrow\,{\w\bmu}_N)|\,,
\\
\label{ratbe711431}
|P^0_{K} ({\w\bmu}_N \rightarrow\,{\w\bmu}_N)| & =
\sum_{|{\bf n}|=K}
P({\bf n})
\det\left( \begin{pmatrix}{n_j} \\ \frac{{n_j}+{\w\mu}_j-{\w\mu}_i}2
\end{pmatrix} \right)_{1
\le i, j \le N}\,.
\end{align}
Equation (\ref{ratbe71143}) defines the number of nests of trajectories of $N$ random turns vicious walkers initially located at all admissible ${\w\bmu_N}$ and returning after $K$ steps to their initial positions. The generating function $e^{\be h (\frac{M}2 - N)} \w{\cal I}_N (\be | {\bf k}_l)$ enables enumeration of nests of closed paths with stays allowed and with a part of initial/final positions pinned (compare with
\eqref{new16}).

\subsection{Diagonally constrained plane partitions}

The representation $\tr_N(e^{\cQ} e^{-\be H})$ (\ref{ratbe7003}) is estimated with the help of (\ref{cor:lin667})
at $M \gg 1$:
\begin{align}
\nonumber
\frac{\tr_N(e^{\cQ} e^{-\be H})}{e^{\be h M/2}}\,=\,
\frac{1}{N!} \int_{{\sf S}^{N}} {\cal P}_{\CK}
(e^{-i{\bf p}_N}, e^{i{\bf p}_N}, {\bf a}_M)\,| \CV (e^{i{\bf p}_N}) |^2 \\
\label{ratbe7113}
\times\,e^{{\be}\sum_{l=1}^N
(\cos{p}_l - h)}
\frac{d^Np}{(2\pi)^N}
\,,
\end{align}
where the integration is defined in (\ref{cor:lin667}), and
${\cal P}_{\CK} (e^{-i{\bf p}_N}, e^{i{\bf p}_N}, {\bf a}_M)$ is given by (\ref{cor:lin552}).

Let us introduce the notation
${\bf a}^\ga_{M} \equiv \log \ga\cdot (1, 2, \ldots, M)$,
which implies that the parametrization $\al_n= n \log \ga$, $0<\ga\le 1$, is used in
${\cal P}_{\CK}$ (\ref{cor:lin552}) (cf. Section~\ref{sec55}). Then we obtain from (\ref{ratbe7113}) the
estimate at $1\ll M \ll \be$:
\begin{align}
\frac{\tr_N(e^{\cQ(\ga)} e^{-\be H})}{e^{\be h M/2}}\,&\simeq\, {\cal P}_{\CK} ({\bf 1}_N, {\bf 1}_N, {\bf a}^\ga_{M})\,
V_N(\be, h)\,,
\label{ratbe072} \\
V_N(\be, h)\,&\equiv\,
\frac{e^{\be N(1-h)}\, \mathfrak{I}_N}{\be^{N^2/2}}\,=\,
e^{\be N(1-h)-\frac{N^2}{2}\log\be + \varphi_N}\,,\quad \varphi_N\equiv \log \mathfrak{I}_N \,,
\label{ratbe791}
\end{align}
where $\mathfrak{I}_N$ in (\ref{ratbe791}) is Mehta integral \cite{meh} expressed in terms of the Barnes $G$-function \cite{barn}:
\begin{equation}
\nonumber
\mathfrak{I}_N\,=\, \frac{G(N+1)}{(2\pi)^{N/2}}\,,
\qquad G(N+1)\,\equiv\,\frac{(N !)^N}{1^1\,2^2\,\ldots N^N }\,=\,
\prod^N_{k=1} \Gamma(k)\,.
\end{equation}
The behaviour of $\mathfrak{I}_N = e^{ \varphi_N}$ is governed at $1\ll N\ll M$ by the estimate \cite{bmumn}:
\begin{equation}
{{\varphi}}_N\,=\,\frac{N^2}{2} \log N\,-\,
\frac{3 N^2}{4}\,+\,{\cal O}(\log
N)\,,\qquad N\gg 1\,. \label{spdfpxx6}
\end{equation}
It is seen from (\ref{spdfpxx6}) that $V_N(\be, h)$ (\ref{ratbe791}) only depends on
$\be$, $N$ via the ratio
$\frac{\be}{N}$ at $\be> N\gg 1$.

The coefficient ${\cal P}_{\CK} ({\bf 1}_N, {\bf 1}_N, {\bf a}^\ga_{M})$ \eqref{ratbe072} arises at $q\to 1$ from (\ref{cor:lin552}) under the $q$-parametrization (\ref{rep21}):
\begin{equation}
<e^{\cQ (\ga)}>_{N,q}\,\equiv\,
{{\cal P}}_{\CK}
\Bigl(\textbf{q}_N, \frac{\textbf{q}_N}{q}, {\bf a}^\ga_{M}\Bigr)
\underset{q\to 1} \longrightarrow\,{\cal P}_{\CK} ({\bf 1}_N, {\bf 1}_N, {\bf a}^\ga_{M})\,,
\label{ratbe721}
\end{equation}
where $<e^{\cQ (\ga)}>_{N,q}$ is defined by (\ref{ratbee707}).
The limiting expression (\ref{ratbe721}) leads from
the generating function of plane partitions with fixed sum of their diagonal parts confined in
$N\times N\times {\CM}$ box to the case of box with infinite height:
\begin{align}
\nonumber
{\cal P}_{\CK}({\bf 1}_N, {\bf 1}_N, {\bf a}^\ga_{M}) )\Big|_{M\gg 1} & =\,\ga^{\frac{N}{2} (N+1)}\,{G}(N, N, {\CM} |\,1, \ga)\Big|_{M\gg 1}
\\
\nonumber
& \underset{M\to\infty} \longrightarrow\, \ga^{\frac{N}{2} (N+1)}
\lim_{q\to 1}\,
\displaystyle{
\prod_{i=1}^{N} \prod_{j=1}^{N} \frac{1}{1- \ga
q^{i+j-1} }}\,.
\end{align}

\noindent $\bullet\,\,$ The mean value of the generating exponential (\ref{new11}) is estimated by means of (\ref{ratbe721}):
\begin{equation}
\label{ratbe723}
G_{N, \be} ({\bf a}^\ga_{M})\Bigl|_{1\ll M\ll\be}\,\simeq\,
\ga^{\frac{N}{2} (N+1)}
\frac{{G}(N, N, {\CM} |\,1, \ga)\Big|_{M\gg 1}}{{A}(N, N, {\CM})\Big|_{M\gg 1} }\,.
\end{equation}
According to (\ref{ratbe794}) and (\ref{cor:llin5591}), the denominator in (\ref{ratbe723}) is the number ${A}(N, N, {\CM})$ of unconstrained plane partitions in $N\times N\times {\cal M}$ box of large height. According to (\ref{cor:lin559}), the generating function ${G}(N, N, {\CM} |\,1, \ga)$ (\ref{ratbe707}) is of a polynomial form. Application of ${\mathcal D}^l_{\al}$ to (\ref{ratbe723}) at $\ga=e^{{\al}/N}$ gives:
\begin{equation}
{{\langle\langle}} {\sf M}^l {{\rangle\rangle}}_{N, \be}\Big|_{1\ll M\ll\be}\,
\simeq\,
\frac{N^{-l} \sum_{\{{\bmu}_N\}} |{\bmu}_N|^l
S_{\bmu_N-\bdl_N} (\mathbf{1}_N) S_{\bmu_N-\bdl_N} (\mathbf{1}_N)
\Big|_{M\gg 1}}{{A}(N, N, {\CM})\Big|_{M\gg 1} }\,.
\label{ratbbbe723}
\end{equation}
The numerator in right-hand side of (\ref{ratbbbe723}) may be viewed as a sum of the terms $(\frac{m}{N})^l {A}(N, N, {\CM} |\,m)$, where
\begin{equation}
\label{new40}
{A}(N, N, {\CM} |\,m) \equiv \sum_{{\bmu}_N \vdash m}
S_{\bmu_N-\bdl_N} (\mathbf{1}_N) S_{\bmu_N-\bdl_N} (\mathbf{1}_N)
\end{equation}
is the number of \textit{diagonally constrained}  plane partitions with $\tr_N{\bpi} = m-\frac{N}{2}(N+1)$ confined in 
$N\times N\times {\cal M}$ box.

\noindent $\bullet\,\,$ The mean value of ${\varPi}_{\bf k}$ is estimated:
\begin{equation}
\label{ratbe724}
{{\langle\langle}} {\varPi}_{\bf k} {{\rangle\rangle}}_{N, \be}
\Bigl|_{1\ll M\ll\be}\,\simeq\,
\frac{{\w {\cal P}}_{\CK}
(\textbf{1}_N, \textbf{1}_N, {\bf k}_l)}{{A}(N, N, {\CM}) }\,,
\end{equation}
where ${\w {\cal P}}_{\CK}
(\textbf{1}_N, \textbf{1}_N, {\bf k}_l)$ is the number of the plane partitions given by (\ref{cor:lenn5571}) at $q\to 1$:
\begin{equation}\label{new99}
\lim_{q\to 1} <{\varPi}_{\bf k} >_{N, q} \,=\, {\w {\cal P}}_{\CK} (\textbf{1}_N, \textbf{1}_N, {\bf k}_l) \,=\,
\sum_{\{\w{\blad}\}}
S_{\w{\blad}} (\textbf{1}_N)
S_{\w{\blad}} (\textbf{1}_N)\,.
\end{equation}
The numbers ${\w {\cal P}}_{\CK}
(\textbf{1}_N, \textbf{1}_N, {\bf k}_l)$ (\ref{new99}) enumerate the plane partitions diagonally constrained in the sense that $l$ columns on the principal diagonal are of prescribed heights and positions (see {\sf Definition~3}).
The denominators in \eqref{ratbbbe723} and (\ref{ratbe724}) are the same.
In the case of ${\bf k}_l= {\bdl}_l$, the estimate (\ref{ratbe724}) is given by ${\w {\cal P}}_{\CK}
(\textbf{1}_N, \textbf{1}_N, {\bdl}_l)$ (\ref{cor:lin5572}).

The number of the diagonally constrained plane partitions ${\w {\cal P}}_{\CK}
(\textbf{1}_N, \textbf{1}_N, {\bf k}_l)$ (\ref{new99}), as well as the number of closed lattice trajectories $\w{\cal P}_{K, N} ({\bf k}_l)$ (\ref{ratbe71143}), both mean summation $\sum_{\{{\w\bmu}_N\}}$, which takes into account  either fixed positions of columns or initial/final positions of walkers are pinned at ${\bf k}_l$.
In turn, the representation  \eqref{new40} looks similar to \eqref{new18} enumerating such nests of closed paths that initial/final positions are given by partitions of certain positive integers.

\section{The generating function $G_\be ({\bf a}_M)$}
\label{sec6011}

\subsection{Determinantal representation}
\label{sec601}

Let us proceed with $G_\be ({\bf a}_M) = \l\l e^{\cQ ({\bf a}_M)} \r\r_{\be}$
(Section~\ref{sec2}) in the case when ${\sf trace}$ includes summation over the Bethe solutions and the numbers of particles \cite{KBI2, col1, col2}. We obtain:
\begin{align}
\Tr\bigl(e^{\cQ}\, e^{-\be H}\bigr) & =\,\sum_{N=0}^{M}
\sum_{\{{\bth}_N\}} \langle e^{\cQ}\, e^{-\be H} \rangle_N
\nonumber
\\
\label{cor:lin669}
& = e^{\be h M/2}\,\Bigl(1+\sum_{N=1}^{M}
\sum_{\{{\bth}_N\}} \det
(e^{- \be {\h\ep}} {e^{\h\al}})\Bigr)\,.
\end{align}
Equation (\ref{cor:lin665}) is used in (\ref{cor:lin669}), and  averaging at $N=0$ is over $\mid\Uparrow\rangle$. Besides,  $Z=\Tr\bigl(e^{-\be H}\bigr)$ (\ref{cor:lin5}) results from (\ref{cor:lin669}) at $\h\al = 0$.

Expression (\ref{cor:lin669}) is calculated by the Laplace formula for determinant of sum of two matrices \cite{KBI2}:
\begin{align}
\label{tr2}
\Tr\bigl(e^{\cQ}\, e^{-\be H} \bigr)\,=\,\frac{e^{\be h M/2}}{2}\sum_{\ell=\pm 1}
\bigl({\sf D}_{+}^{(\ell)}({\h\al})+\ell\,
{\sf D}_{-}^{(\ell)}({\h\al}) \bigr)\,,
\\
\label{tr1}
{\sf D}_{\pm}^{(\ell)}({\h\al}) \equiv \det({\h I}+ \ell e^{- \be {\h\ep}} {e^{\h\al}})_{p\in {\sf S}^\pm}
\,, \qquad \ell=\pm 1\,,
\end{align}
where the subscript $p\in {\sf S}^\pm$ reminds that $M\times M$ matrices $e^{\h\al}$ and $e^{- \be {\h\ep}}$ are parameterized by elements of ${\sf S}^\pm$ (\ref{qanal27777}) (cf.~(\ref{cor:lin55311}) and (\ref{mpcf1}); for instance, $\h\ep\equiv {\diag} \{\ep_p\}_{p\in {\sf S}^\pm}$, where $\ep_p\equiv \ep (p)$ is given by (\ref{egen})), and $\h I$ is unit $M\times M$ matrix.

Further, let us consider the determinantal identities:
\begin{equation}
\label{cor:lin674}
{\sf D}_{\pm}^{(\ell)} ({\h\al}) =
{\sf G}_{\pm}^{(\ell)}\,
{\sf D}_{\pm}^{(\ell)}({\h\al=0})\,,\quad
{\sf G}_{\pm}^{(\ell)} \equiv \det({\h I}\,+ {\h M}_{\rm xx}^{(\ell)})_{p\in {\sf S}^\pm}\,,
\end{equation}
where
\begin{equation}
{\h M}_{\rm xx}^{(\ell)}\,\equiv\, ({{e^{\h\al}}}-{\h I}){\h f}^{(\ell)}\,,\qquad
{\h f}^{(\ell)} \equiv
({\h I} + \ell\,e^{\be
{\h\ep}})^{\1}\,.
\label{cor:lin67002}
\end{equation}
The determinantal representation for $G_\be ({\bf a}_M)$ (\ref{cor:lin5}) resulting from (\ref{tr2}) and
(\ref{cor:lin674}) is reduced under the conventional choice of ${\bf a}_M$ to $\l\l e^{\al Q(m)} \r\r_\be$ \cite{col2}.

Using (\ref{ratbe7003}) and (\ref{cor:lin669}), one arrives at the following

\vskip0.3cm \noindent
\noindent{\bf Statement:\,}

\noindent\textit{$\bullet\,\,$
Total trace of the Boltzmann-weighted generating exponential is represented at large enough $M$}:
\begin{equation}
\label{ratbe7103}
\Tr\bigl(e^{\cQ}\, e^{-\be H} \bigr)\,\simeq\, e^{\be h M/2} \Bigl(1+ \sum_{N=1}^{M\gg 1} e^{-\be h N} {\cal I}_N (\be, {\bf a}_{M})\Bigr)\,,
\end{equation}
\textit{where ${\cal I}_N (\be, {\bf a}_{M})$ is given by} (\ref{ratbe711432}).

\vskip0.3cm
\noindent\textit{$\bullet\,\,$ The correlation function of ${\varPi}_{\bf k}$ defined by \eqref{cor:lin611}, \eqref{cor:lin61} has the form, with regard at \eqref{ratbe711433} and \eqref{ratbe7103}}:
\begin{equation}
\l\l {\varPi}_{\bf k} \r\r_\be = \frac{\varPhi (\be, h, {\bf k}_l)}{Z}\,,
\label{ratbe711437}
\end{equation}
\textit{where}
\begin{equation}
\varPhi (\be, h, {\bf k}_l) \simeq
e^{\be hM/2} \sum_{N=l}^{M\gg 1}
e^{-\be h N} \w{\cal I}_N (\be | {\bf k}_l)\,,
\label{ratbe7114}
\end{equation}
\textit{and $Z=\Tr\bigl(e^{-\be H}\bigr)$ \eqref{cor:lin5} arises from \eqref{ratbe7103} provided that
${\bf a}_M$ consists of zeros}.
\vskip0.2cm

Therefore, $\varPhi (\be, h, {\bf k}_l)$ (\ref{ratbe7114}) is the
``superposition''
of the generating functions $e^{\be h (\frac{M}2 - N)} \w{\cal I}_N (\be | {\bf k}_l)$ of nests of closed paths of random turns
vicious walkers such that $l$ initial/final positions are pinned and stays of all walkers are allowed:
\begin{equation}
\mathcal D^K_{\be/2} \varPhi (\be, h, {\bf k}_l) \equiv \sum_{N=l}^{M \gg 1}
\sum_{p=0}^{K}
\begin{pmatrix}
K\\
p
\end{pmatrix}
\bigl(h(M-2N) \bigr)^p\,
\w{\cal P}_{K-p, N} ({\bf k}_l)\,,
\label{ratbe711439}
\end{equation}
where $\w{\cal P}_{K-p, N} ({\bf k}_l)$ are defined by \eqref{new14} and (\ref{ratbe71143}).

\subsection{Differentiation of $G_\be ({\bf a}_M)$}

Let us consider differentiation of the generating function $G_\be ({\bf a}_M)$. We introduce the shortening notations ${\sf G} \equiv {\sf G}_{\pm}^{(\ell)}$, ${\h R}\equiv (\h I +{\h M}_{\rm xx}^{(\ell)})^{\1}$, and obtain  the first order derivative:
\begin{equation}
{\sf G}^{\1}\cd_{k_1} {\sf G}\,=\, e^{\al_{k_1}} \tr\big({\h {R}}\, \h\dl_{k_1} {\h f} \big)\,=\,
e^{\al_{k_1}}
{\bar R}_{k_1, k_1} \,,
\label{cor:bin2}
\end{equation}
where $\cd_l\equiv\cd/\cd\al_{l}$ and ${\h\dl}_{l}\equiv \cd_{l} {\h\al}$. The diagonal matrix ${\h f}\equiv {\h f}^{(\ell)}$ is given by (\ref{cor:lin67002}), and ${\bar R}_{k_1, k_1}$ is the diagonal entry of the matrix
${\bar R}\equiv \{{\bar R}_{m n}\}_{1\le m, n\le M}$, where
\begin{equation}
{\bar R}_{m n}\,\equiv\, \frac1M \sum_{p, q \in {\sf S}^{\pm}} f_p\,{R}_{p q} e^{i( m q-np)} \,,
\label{cor:bin4}
\end{equation}
and ${\sf S}^{\pm}$ are given in (\ref{qanal27777}).
The second order derivative
of ${\sf G}$ is obtained,
\begin{eqnarray}
&& {\sf G}^{\1} \cd^2_{k_1, k_2} {\sf G}\,=\,e^{\al_{k_1}+ \al_{k_2}}
\left|\,\begin{matrix} {\bar R}_{k_1, k_1} & {\bar R}_{k_1, k_2}
\\ {\bar R}_{k_2, k_1} & {\bar R}_{k_2, k_2} \end{matrix}\, \right|\,,
\label{cor:bin5}
\end{eqnarray}
since $\tr\big({\h
R}\,\h \dl_{k_1} {\h f} {\h
R}\,\h\dl_{k_2} {\h f} \big)$
takes the form ${\bar R}_{k_1, k_2} {\bar R}_{k_2, k_1}$ due to (\ref{cor:bin4}).

With regard at (\ref{cor:bin2}) and (\ref{cor:bin5}), one obtains the following

\noindent{\bf Proposition~9:} \textit{The function ${\sf G}$ defined by \eqref{cor:lin674} is the generating function of the minors of the matrix} ${\bar R}$  (\ref{cor:bin4}),
\begin{equation}
{\sf G}^{\1} \cd^{\,l}_{k_1, k_2,\dots ,
k_{l}} {\sf G}\,=\, e^{\al_{k_1} + \al_{k_2} + \ldots + \al_{k_l}}\,{\det}_l {\sf R}\,,
\label{cor:bin6}
\end{equation}
\textit{where $\cd^{\,l}_{k_1, k_2,\dots ,
k_{l}}$ is defined by \eqref{cor:lin61}, $M \ge k_1 > k_2 > \cdots > k_l \ge 1$, and ${\det}_l
{\sf R}$ is the minor corresponding to
submatrix of $l^{\rm th}$ order
$\{{\sf R}_{i j}\}_{1 \le i, j  \le l} \equiv \{{\bar R}_{k_i, k_j}\}_{1 \le i, j \le l}$}.

\noindent{\bf Proof:} We use induction with the base case (\ref{cor:bin2}) and induction step consisting in validity of (\ref{cor:bin6}) at $l\mapsto l-1$,
\begin{equation}
{\sf G}^{\1} \cd^{\,l-1}_{k_1, k_2,\dots ,
k_{l-1}} {\sf G}\,=\,e^{\al_{k_1} + \al_{k_2}+\ldots + \al_{k_{l-1}}}\, {\det}_{l-1} {\sf R} \,.
\label{cor:bin8}
\end{equation}
Then, the relation
$\cd_{\al_{k_n}} {{\bar R}}_{k_i, k_j}\,=\,e^{\al_{k_n}} {{\bar R}}_{k_i, k_n}
{{\bar R}}_{k_n, k_j}$ leads from (\ref{cor:bin8}) to
\begin{equation}
\cd^{\,l}_{k_1, k_2,\dots ,
k_{l-1}, k_{l}} {\sf G}\, = \,e^{\al_{k_1} + \al_{k_2}+ \ldots + \al_{k_{l-1}}}\, \bigl({\sf G}\, \cd_{k_{l}} {\det}_{l-1} {\sf R}\,+\,{\det}_{l-1} {\sf R}\, \cd_{k_{l}} {\sf G}\bigr)\,.
\label{cor:bin9}
\end{equation}
The derivative of ${\det}_{l-1} {\sf R}$ is given by:
\begin{equation}
\cd_{k_{l}}{\det}_{l-1} {\sf R}\,=\,\sum \limits^{l-1}_{i=1} (-1)^{l+i} {\sf R}_{l i}\,
\left|\,\begin{matrix} {\sf R}_{1 1} &
{\sf R}_{1 2} & \dots &
\check {\sf R}_{1 i} & \dots & {\sf R}_{1 l}
\\ {\sf R}_{2 1} & {\sf R}_{2 2} & \dots &
\check {\sf R}_{2 i} & \dots & {\sf R}_{2 l}\\
\dots & \dots & \dots &\dots &\dots &\dots \\
{\sf R}_{l-1, 1} & {\sf R}_{l-1, 2} & \dots &
\check {\sf R}_{l-1, i} & \dots & {\sf R}_{l-1, l}
 \end{matrix}\, \right|\,,
\label{cor:bin10}
\end{equation}
where $\check {\sf R}_{i j}$ implies that the column is omitted. The statement (\ref{cor:bin6})
arises from (\ref{cor:bin9})
due to (\ref{cor:bin2}) and (\ref{cor:bin10}). $\Box$

\noindent $\bullet\,$ It is seen from (\ref{tr2}), (\ref{tr1}) that $G_\be ({\bf a}_M)$ at large $M$ is due to $\ell=+1$
whereas the terms at $\ell=-1$ are mutually cancelled since ${p\in {\sf S}^\pm}$ is replaced by ${p\in {\sf S}}$. {\sf Proposition~9} demonstrates that (\ref{cor:lin611}) on infinite chain takes the determinantal form since ${\sf G}$ tends to unity at ${{\textbf a}_{_M} \to 0}$:
\begin{align}
\nonumber
\l\l {\varPi}_{\bf k} \r\r_{\be} \Bigr|_{M\to\infty}  &=
\frac{\varPhi (\be, h, {\bf k}_l)}{Z} \Bigr|_{M\to\infty} \\
&= \lim_{M\to\infty} \lim\limits_{{\textbf a}_{_M} \to 0} {\cd}^l_{\al_{k_1} \al_{k_2} \ldots \al_{k_l}}
{\sf G} = {\det} \bigl(
{f}_{k_i, k_j} \bigr)_{1 \le i, j \le l}\,.
\label{cor:bin11}
\end{align}
The entries ${f}_{k_i, k_j}$ are given by
(\ref{cor:bin4}),
where integration over ${p\in {\sf S}}$ replaces summation
over ${p\in {\sf S}^\pm}$, and $\h R$ tends to unit matrix. Equation \eqref{cor:bin11} is of approximate meaning at $M\gg 1$.

\section{Discussion}
\label{sec6}

The enumerative combinatorial implications of the matrix elements of the generating exponential and of appropriate distributions of flipped spins are in the focus of the present paper. The Bethe state-vectors are taken in terms of the Schur polynomials, and the Heisenberg $XX$ chain is considered for concreteness.
Connections are elaborated between the mean values of the generating exponential, the random walks over closed trajectories with constrained initial/final positions, and the plane partitions with constrained diagonal parts. The number of nests of paths of vicious walkers is expressed in terms of the lacunary sums of the binomial coefficients, and generalized Ramus's identities are derived.

More specifically, the transition amplitude between $N$-particle states is studied as the generating function of nests of random turns walks.
A relationship between the entries of powers of the circulant matrix,
the lacunary sums of the binomial coefficients, and non-intersecting walks of vicious walkers is unraveled by means of Ramus's identity and its generalizations. When the length of the chain is large enough, the number of nests of paths is connected with the Gross-Witten partition function, as well as with the problem of enumeration of increasing subsequences of random permutations.

The determinantal representation for the norm-trace generating function of boxed plane partitions with fixed height of diagonal parts is obtained
as form-factor of the generating exponential over off-shell $N$-particle Bethe states.

Two trace definitions are admitted to calculate the Boltzmann-weighted mean values: the trace over $N$-particle on-shell Bethe states and the total trace including summation over numbers of flipped spins. The matrix elements are considered for powers of the first moment of the distribution of flipped spins, as well as for inconsecutive flipped spins.

The $N$-particle matrix elements of powers of the first moment of the distribution of flipped spins
as well as of sequence of flipped spins are expressed at long enough chain through the generating function of nests of random turns walks. Closed trajectories of random turns vicious walkers are characterized by constrained initial/final positions: for each nest its initial/final coordinates either are partitions of certain integers or coincide partially with the positions of flipped spins.

In the limit when the evolution parameter grows faster than the length of the chain, the trace is related to enumeration of plane partitions with constrained diagonal parts.
The estimates demonstrate a similarity with the case of long chain: the diagonally constrained plane partitions are characterized either by fixed total traces
or diagonal parts of fixed height are labelled by the flipped spins positions.

In the case of the total trace, the mean value of projector of inconsecutive flipped spins acquires for infinite chain the determinantal representation interpreted in terms of nests of paths of random turns walks such that stays of all walkers are allowed and a subset of initial/final points is given by positions of flipped spins.

The results obtained look stimulating from the viewpoint
of further investigation of
the generating functions of constrained plane partitions
in framework of the vertex models, of the phase model, and of the $XY$ model (cf.~\cite{gauss}) along the lines presented.

\vskip-0.3cm
\section*{{\small Acknowledgement}}
\vskip-0.3cm
{\small This work was supported by the Russian Science Foundation (Grant 18-11-00297).}

\section*{Appendix I}

In order to derive (\ref{qanal272}), let us first
express the number of nests of paths of random turns vicious walkers using (\ref{qanal13}):
\begin{equation}
\mathfrak{G}^{\,0}({\bmu^L}; {\bmu^R} | K) = \sum_{|{\bf n}|=K} P({\bf n})\,
{\bold\Delta}^{\bf n}_{{\bmu^L}; {\bmu^R}}\,,
\nonumber
\eqno({\rm AI.1})
\end{equation}
where ${\bf n}=(n_1, n_2, \ldots, n_N)$, $|{\bf n}|\equiv n_1+n_2+ \ldots + n_N$, and $P({\bf n})$ is the multinomial coefficient (\ref{qanal271}). Further,
\begin{equation}
{\bold\Delta}^{\bf n}_{{\bmu^L}; {\bmu^R}} \equiv \prod_{j=1}^{N}  ({\bold\Delta}^{n_j} )_{\mu^L_{j}; \mu^R_j} \,,
\nonumber
\eqno({\rm AI.2})
\end{equation}
where
$({\bold \Delta}^{n})_{j m}$ is defined by \eqref{avv},
and $({\bold \Delta}^{0})_{j m} = {\delta}_{j m}$.
The commutation relation
\begin{equation}
[ H_{\rm xx}, \si_{l_1}^{-} \si_{l_2}^{-} \dots \si_{l_N}^{-}] = \sum_{k=1}^N \si_{l_1}^{-} \dots \si_{l_{k-1}}^{-}\, [ H_{\rm xx}, \si_{l_k}^{-}]\, \si_{l_{k+1}}^{-} \dots\si_{l_N}^{-}
\nonumber
\end{equation}
supplied with the relations $H_{\rm xx} \mid \Uparrow\rangle = 0$ and $\si^z_k \mid \Uparrow\rangle =\mid \Uparrow\rangle$
enables us to obtain (AI.1) at $K=1$ as the base case of induction. As induction step, it is assumed that (AI.1) is valid at $K-1$. We put $H_{\rm xx}^{K}= H_{\rm xx}^{K-1} H_{\rm xx}$ in (\ref{qanal13}) to prove (AI.1) and obtain:
$$
\begin{array}{rcl}
\displaystyle{
\mathfrak{G}^{\,0}({\bmu^L}; {\bmu^R} | K)} &=& \displaystyle{
\sum_{l=1}^{N}\sum_{k=1}^M
{\bold\Delta}_{\mu^R_l,\,k}
\sum_{|{\bf n}|=K-1} P({\bf n})\,
{\bold\Delta}^{\bf n}_{{\bmu^L}; {(\mu^R_1, \ldots, \mu^R_{l-1}, k, \mu^R_{l+1}, \ldots, \mu^R_N) }}}
\\
&=&\displaystyle{
\sum_{l=1}^{N} \sum_{|{\bf n}|=K-1} P({\bf n})\, {\bold\Delta}^{{\bf n}+{\bf e}_{l}}_{{{\bmu}^L}; {{\bmu^R}}}
}\,,
\end{array}
\eqno({\rm AI.3})
$$
where $N$-tuples ${\bf e}_{l}$ are defined in (\ref{ratbe7272}).
The multinomial theorem demonstrates that (AI.3) leads to (AI.1). The determinantal generalization of (AI.1) leads to
(\ref{qanal272}), where the non-intersection requirements
are taken into account.

Now let us turn to the verification of (\ref{qanal272}). The corresponding difference equation (\ref{mpcf6}) is a tool of verification rather than derivation of (\ref{qanal272}), as stressed in \cite{forr2}.
Equation (\ref{qanal272}) is reduced at $K=0$ to the orthogonality (\ref{tuc3}) and conjectured at arbitrary $K$ due to validity of (\ref{qanal13}). Here we shall verify that (\ref{qanal272}) indeed obeys (\ref{mpcf6}) ($h=0$) as the generalization of (\ref{dmcf}) at $N>1$.

Induction with respect to $N$ enables to prove Eq.~(\ref{qanal272}) provided that the base case $N=1$
is given by (\ref{avv})
and (\ref{dmcf}). The induction step is to assume that (\ref{qanal272}) fulfills
(\ref{mpcf6}) ($h=0$), where the partitions $\bmu^L$ and $\bmu^R$ are of length $N-1$
so that the minors in
(\ref{qanal272}) are of the size $(N-1)\times(N-1)$.
The proof is based on the identity:
\begin{equation}
\mathfrak{G}^{\,0} ({\bmu^L}; {\bmu^R}\,| K+1) \, =\,\sum_{p=0}^{K+1} \begin{pmatrix}
K+1\\
p
\end{pmatrix} \sum_{m=1}^{N} (-1)^{N+m}\,\mathfrak{G}(K+1-p,p,m) \,,
\nonumber
\eqno({\rm AI.4})
\end{equation}
where $\bmu^L$ and $\bmu^R$ are of length $N$. The shortening notation is introduced in (AI.4):
\begin{equation}
\mathfrak{G} (K,P,m)
\equiv \mathfrak{G}^{\,0} ({\bmu^L_{N-1}}(m); {\bmu^R_{N-1}}\,| K)\,\mathfrak{G}^{\,0} ({\mu^L_m}; {\mu^R_N}\,| P)\,,
\nonumber
\eqno({\rm AI.5})
\end{equation}
where ${\bmu^L_{N-1}}(m) \equiv (\mu^L_1, \mu^L_2, \ldots, \mu^L_{m-1}, \mu^L_{m+1}, \ldots, \mu^L_N)$ and $\bmu^R_{N-1}$ are of length $N-1$, and $K$, $P$, $m$ are non-negative integers. The identity (AI.4) is due to expanding the determinant in (\ref{qanal272}) along $N^{\rm th}$ column.

The use of (AI.4)
in right-hand side of (\ref{mpcf6}) (after $K+1 \mapsto K$) gives two identities (labelled by the signs $\pm$):
\begin{equation}
\sum_{k=1}^{N}\mathfrak{G}^{\,0} ({\bmu^L}; {\bmu^R}\pm {\bf e}_k\,| K)\, =\,
\sum_{p=0}^{K} \begin{pmatrix}
K\\
p
\end{pmatrix} \sum_{m=1}^{N} (-1)^{N+m} (\mathfrak{G}^{\,\pm 1}_{p,m} + \mathfrak{G}^{\,\pm {\bf e} }_{p,m}) \,,
\nonumber
\eqno({\rm AI.6})
\end{equation}
where
\begin{align}
\nonumber
\mathfrak{G}^{\,\pm {\bf e} }_{p,m} & \equiv \mathfrak{G}^{\,0} ({\mu^L_m}; {\mu^R_N}\,| p) \sum_{k=1}^{N-1} \mathfrak{G}^{\,0} ({\bmu^L_{N-1}}(m); {\bmu^R_{N-1}}\pm {\bf e}_k\,| K-p) \,,\\
\nonumber
\mathfrak{G}^{\,\pm 1}_{p,m} & \equiv \mathfrak{G}^{\,0} ({\mu^L_m}; {\mu^R_N}\pm 1\,| p)\,\mathfrak{G}^{\,0} ({\bmu^L_{N-1}}(m); {\bmu^R_{N-1}}\,| K-p)\,.
\end{align}
In turn, the
series (AI.4) itself is represented as
\begin{align}
\mathfrak{G}^{\,0} ({\bmu^L}; {\bmu^R}\,| K+1) &  =\,\sum_{m=1}^{N} (-1)^{N+m}
\mathfrak{G} (K+1, 0, m)
\nonumber
&&({\rm AI.7})
\\&
+\,\sum_{p=1}^{K} \begin{pmatrix}
K+1\\
p
\end{pmatrix}
\sum_{m=1}^{N} (-1)^{N+m} \mathfrak{G} (K+1-p,p,m)
\nonumber
&&({\rm AI.8})
\\
& +\,\sum_{m=1}^{N} (-1)^{N+m}
\mathfrak{G} (0, K+1, m)\,,
\nonumber
&& ({\rm AI.9})
\end{align}
where the notation ({\rm AI.5}) is used. The representation  ({\rm AI.7}), ({\rm AI.8}),  ({\rm AI.9})
is compared with right-hand side of (AI.6) so that ({\rm AI.7}) is matched to
the contribution at $p=0$ given by
$\mathfrak{G}^{\,\pm {\bf e} }_{0,m}$ inside the brackets, and ({\rm AI.9}) is matched to the contribution at $p=K$ given by
$\mathfrak{G}^{\,\pm 1 }_{K,m}$. Further, the Pascal relation
\begin{equation}
\begin{pmatrix}
K+1\\
p
\end{pmatrix}\,=\,\begin{pmatrix}
K\\
p-1
\end{pmatrix}\,+\,
\begin{pmatrix}
K\\
p
\end{pmatrix}
\nonumber
\eqno({\rm AI.10})
\end{equation}
is used in the line ({\rm AI.8}). The base case is applied to $\mathfrak{G}^{\,0} ({\mu^L_m}; {\mu^R_N}\,| p)$ in the contribution due to the first term in ({\rm AI.10}), whereas the induction assumption
is applied to $\mathfrak{G}^{\,0} ({\bmu^L_{N-1}}(m); {\bmu^R_{N-1}}\,| K+1-p)$ in the contribution due to the second term in ({\rm AI.10}).
Eventually, $\mathfrak{G}^{\,0} ({\bmu^L}; {\bmu^R}\,| K+1)$ coincides with right-hand side of the sum of two identities (AI.6).

\section*{Appendix II}

It is straightforward to obtain
useful identities provided that $\bigl({\bold\Delta}^K \bigr)_{j m}$ given by \textsf{Proposition~4}, on one hand,
and by \cite{rim1, rim2}, on another, are equated each to other. Without reproducing the appropriate formulae from \cite{rim1, rim2}, we simply specify, according to \textsf{Proposition~4}, the matrix $\bigl({\bold\Delta}^K \bigr)_{j m}\equiv \bigl({\bold\Delta}^K \bigr)_{j - m}$ of the size $6\times 6$ to $K=14$:
\begin{equation}
\begin{array}{l}
\bigl({\bold\Delta}^{14} \bigr)_{0} = \begin{pmatrix} 14 \\ 1 \end{pmatrix}_{3} = 5462\,,\\ [0.3cm] \bigl({\bold\Delta}^{14} \bigr)_{2} = \begin{pmatrix} 14 \\ 0 \end{pmatrix}_{3} =  \bigl({\bold\Delta}^{14} \bigr)_{4} = \begin{pmatrix} 14 \\ 2 \end{pmatrix}_{3} = 5461\,,\\[0.3cm]
\bigl({\bold\Delta}^{14} \bigr)_{1} =
\bigl({\bold\Delta}^{14} \bigr)_{3} = \bigl({\bold\Delta}^{14} \bigr)_{5} = 0 \,.
\end{array}
\nonumber
\end{equation}
We obtain in notations \cite{rim1, rim2}:
\begin{equation}
\begin{array}{l}
\bigl({\bold\Delta}^{14} \bigr)_{0} =
\displaystyle{\frac{a_1}{6} = \frac{2(2^{14}+2)}{6}}\,,\\[0.3cm]
\bigl({\bold\Delta}^{14} \bigr)_{2} = \bigl({\bold\Delta}^{14} \bigr)_{4} = \displaystyle{ \frac{a_3}{6}=
 \frac{2(2^{14}-1)}{6}
 = \frac{a_1}{6} - 1} \,.
\end{array}
\nonumber
\end{equation}

Further, we specify  $\bigl({\bold\Delta}^K \bigr)_{j - m}$ to $M=6$ and $K=15$:
\begin{equation}
\begin{array}{l}
\bigl({\bold\Delta}^{15} \bigr)_{1} = \begin{pmatrix} 15 \\ 1 \end{pmatrix}_{3} =  \bigl({\bold\Delta}^{15} \bigr)_{5} = \begin{pmatrix} 15 \\ 2 \end{pmatrix}_{3} = 10923
\,,\\ [0.3cm]
\bigl({\bold\Delta}^{15} \bigr)_{3} = \begin{pmatrix} 15 \\ 0 \end{pmatrix}_{3} = 10922
\,,\\[0.3cm]
\bigl({\bold\Delta}^{15} \bigr)_{0} =
\bigl({\bold\Delta}^{15} \bigr)_{2} = \bigl({\bold\Delta}^{15} \bigr)_{4} = 0 \,.
\end{array}
\nonumber
\end{equation}
As well,
\[
\bigl({\bold\Delta}^{15} \bigr)_{1}
 = \bigl({\bold\Delta}^{14}
\bigr)_{0} + \bigl({\bold\Delta}^{14} \bigr)_{2}\,, \qquad \bigl({\bold\Delta}^{15} \bigr)_{3}
= \bigl({\bold\Delta}^{14} \bigr)_{2} + \bigl({\bold\Delta}^{14} \bigr)_{4}\,.\]
We obtain in notations \cite{rim1, rim2}:
\begin{equation}
\begin{array}{l}
\bigl({\bold\Delta}^{15} \bigr)_{1} = \bigl({\bold\Delta}^{15} \bigr)_{5} = \displaystyle{ \frac{a_2}{6}=
\frac{2(2^{15}+1)}{6}}\,,\\[0.3cm]
\bigl({\bold\Delta}^{15} \bigr)_{3} =
\displaystyle{\frac{a_4}{6} = \frac{2(2^{15}-2)}{6}
 = \frac{a_2}{6} - 1} \,.
\end{array}
\nonumber
\end{equation}

\end{document}